\documentclass[11pt]{article}
\usepackage{amsmath, amssymb, amsthm, float, fullpage, graphicx, parskip, subcaption, multirow, setspace}
\usepackage{tikz}
\usetikzlibrary{shapes.geometric, positioning}
\usetikzlibrary{quotes, angles}
\usepackage{hyperref}
\usepackage{natbib}
\usepackage{authblk}

\defcitealias{OfficalBaseballRules2013}{Officie Baseball Rules}

\definecolor{PennRed}{RGB}{152, 30 50}
\definecolor{PennBlue}{RGB}{0, 44, 119}
\definecolor{PennGreen}{RGB}{94, 179,70}
\definecolor{PennViolet}{RGB}{141, 76, 145}
\definecolor{PennSkyBlue}{RGB}{14, 118, 188}
\definecolor{PennOrange}{RGB}{243, 117, 58}
\definecolor{PennBrightRed}{RGB}{223,82, 78}

\hypersetup{pdfborder = {0 0 0.5 [3 3]}, colorlinks = true, linkcolor = PennBrightRed, citecolor = PennSkyBlue}

\title{A Hierarchical Bayesian Model of Pitch Framing}
\author[]{Sameer K. Deshpande}
\author[]{Abraham J. Wyner}

\affil[]{The Wharton School \\ University of Pennsylvania}
\date{9 September 2017}

\begin{document}
\maketitle
\def\C{\mathbb{C}}
\def\R{\mathbb{R}}
\def\Q{\mathbb{Q}}
\def\Z{\mathbb{Z}}
\def\N{\mathbb{N}}

\def\P{\mathbb{P}}
\def\E{\mathbb{E}}

\begin{abstract}
Since the advent of high-resolution pitch tracking data (PITCHf/x), many in the sabermetrics community have attempted to quantify a Major League Baseball catcher's ability to ``frame'' a pitch (i.e. increase the chance that a pitch is called as a strike).
Especially in the last three years, there has been an explosion of interest in the ``art of pitch framing'' in the popular press as well as signs that teams are considering framing when making roster decisions.

We introduce a Bayesian hierarchical model to estimate each umpire's probability of calling a strike, adjusting for pitch participants, pitch location, and contextual information like the count. 
Using our model, we can estimate each catcher's effect on an umpire's chance of calling a strike.
We are then able to translate these estimated effects into average runs saved across a season.
We also introduce a new metric, analogous to Jensen, Shirley, and Wyner's Spatially Aggregate Fielding Evaluation metric, which provides a more honest assessment of the impact of framing.
\end{abstract}

\doublespacing
\section{Introduction}
\label{sec:introduction}

\begin{figure}[!h]
\centering
\includegraphics{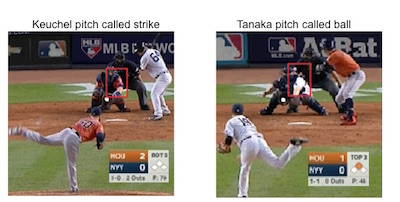}
\caption{Both pitches missed the strike zone (outlined in red) and by rule, should have been called balls. Keuchel's pitch (left) was called a strike while Tanaka's pitch (right) was called a ball. Screenshot source: http://www.fangraphs.com/blogs/how-the-astros-wound-up-with-a-bigger-zone/}
\label{fig:keuchel_tanaka}

\end{figure}

The New York Yankees and Houston Astros played each other in the American League Wild Card game in October 2015, with the winner continuing to the next round of the Major League Baseball playoffs. 
During and immediately after the game, several Yankees fans took to social media expressing frustration that home plate umpire Eric Cooper was not calling balls and strikes consistently for both teams, thereby putting the Yankees at a marked disadvantage.
Even players in the game took exception to Cooper's decision making: after striking out, Yankees catcher Brian McCann argued with Cooper that he was calling strikes on similar pitches when the Astros were pitching but balls when the Yankees were pitching. 
Figure~\ref{fig:keuchel_tanaka} shows two pitches thrown during the game, one by Astros pitcher Dallas Keuchel and the other by Yankees pitcher Masahiro Tanaka.

%\begin{figure}[!h]
%\begin{minipage}{0.1\textwidth}
%\centering
%\includegraphics{keuchel_tanaka.pdf}
%\end{minipage}
%\caption{A}

%\end{figure}

Both pitches were thrown in roughly similar locations, near the bottom-left corner of the \textit{strike zone}, the rectangular region of home plate shown in the figure.
According to the official rules, if any part of the pitched ball passes through the strike zone, the umpire ought to call it a strike.
Keuchel's pitch barely missed the strike zone while Tanaka's missed by a few inches.
As a result, the umpire Cooper ought to have called both pitches a ball. 
That Cooper did not adhere strictly to the official rules is hardly surprising; previous research has shown umpires' ball/strike decisions may be influenced by the race or ethnicity of the pitcher \citep[see, e.g.,][]{Parsons2011, TainskyMillsWinfree2015}, player status as measured by age or ability \citep[see, e.g.,][]{KimKing2014, Mills2014}, and their previous calls \citep{ChenMoskowitzShue2016}.
During the television broadcast of the game, the announcers speculated that the difference in Cooper's strike zone enforcement was the ability of Astros catcher Jason Castro to ``frame'' pitches, catching them in such a way to increases Cooper's chance of calling a strike \citep{Sullivan2015}.

Though pitch framing has received attention from the sabermetrics community since 2008, it has generated tremendous interest in the popular press 
\citep[see, e.g.,][]{Lindbergh2013, Pavlidis2014, Sanchez2015} 
and among team officials \citep[see, e.g.,][]{Drellich2014, Holt2014} in the last three or four years, due to its apparently large impact on team success.
According to \citet{Woodrum2014}, most studies of framing, including the most recent by \citet{JudgePavlidisBrooks2015} for the website Baseball Prospectus, estimate that a good framer can, on average, save his team as many as 25 runs more than the average catcher, over the course of the season.
By the traditional heuristic of 10 average runs per win \citep{Cameron2008}, these results suggest that the way a good framer catches a few pitches per game may be worth as many as an additional 2 to 3 wins, relative to the average catcher. 
%By the traditional heuristic of 10 runs per win \citep{Cameron2008}, these results suggest that the way a good framer catches a few pitches per game may be worth as many as 2 to 3 wins for his team, on par with the offensive production of the game's elite batters.
Despite the ostensibly large impact framing may have on team success, framing itself has been overlooked and undervalued until just recently.
\citet{Sanchez2015} highlights the catcher Jonathan Lucroy, whose framing accounted for about 2 wins in the 2014 and worth about \$14M, writing that ``the most impactful player in baseball today is the game's 17$^{th}$ highest-paid catcher.''

Returning to the two pitches in Figure~\ref{fig:keuchel_tanaka}, Cooper may have been more likely to call the Keuchel pitch a strike because of Castro's framing. 
However, looking carefully at Figure~\ref{fig:keuchel_tanaka}, we see that the two pitches are quite different, making it difficult to immediately attribute the difference in calls to Castro.
First, Keuchel's pitch is much closer to the displayed strike zone than Tanaka's and it was thrown in a 1 -- 0 count while Tanaka's was thrown in a 1 -- 1 count.
%\footnote{The count refers to the number of balls and strikes in a plate appearance. The first number is the number of balls and the second is the number of strikes.}.
We also note that the batters, catchers, and pitchers involved in each pitch are, necessarily, different.
In particular, Keuchel is a left-handed pitcher and Tanaka is a right-handed pitcher. 
Any of these factors may have contributed to Cooper being more likely to call Keuchel's pitch a strike.
Of course, it could also be the case that Cooper was equally likely to call both pitches a strike and the different calls are simply due to noise.
This raises questions: what effect did Castro have on Cooper's called strike probability, over and above factors like the pitch location, count, and the other pitch participants? 
And what impact does such an effect have on his team's success?

Existing attempts to answer these questions fall broadly into two categories: those that do not fit statistical models of the called strike probability and those that do.
The first systematic study of framing \citep{Turkenkopf2008} falls into the former category. 
For each catcher, he counts the number of strikes called on pitches thrown outside an approximate strike zone introduced by  \citet{Walsh2007}. %who approximated the strike zone as a rectangular box over home plate and set the horizontal (resp. vertical) boundaries by first estimating the proportion of called strikes as a function of the pitches horizontal (resp. vertical) coordinates and then identifying the 50\% point.
\citet{Turkenkopf2008} then took the counts of ``extra strikes'' received by each catcher and converted them into a measure of runs saved using his own valuation of 0.16 runs saved per strike. %% where did this number come from? It comes from Turkenkopf's piece. 
Missing from this analysis, however, is any consideration of the other players and the umpire involved in the pitch, as well as the context in which the pitch was thrown (e.g. count, run differential, inning, etc.).
This omission could overstate the apparent impact of framing since it is not immediately clear that a catcher deserves all of the credit for an extra strike.
More recently, \citet{RosalesSpratt2015} proposed an iterative method to distribute credit for a called strike among the batter, catcher, pitcher, and umpire.
Unfortunately, many aspects of their model remain proprietary and thus, the statistical properties of their procedure are unknown.

The second broad category of framing studies begins by fitting a statistical model of the called strike probability that accounts for the above factors.
Armed with such a model, one then estimates the predicted called strike probability with and without the catcher.
The difference in these probabilities reflects the catcher's apparent framing effect on that pitch.
One then estimates the impact of framing by weighting these effects by the value of ``stealing a strike'' and summing over all pitches caught by a catcher.
%By weighting these effects by an appropriate estimate of the value of ``stealing a strike,'' and summing over all pitches caught by a catcher, one can arrive at an estimate of the value of framing. 
\citet{Marchi2011} fit a mixed-effects logistic regression model, expressing the log-odds of a called strike as a function of the identities of the pitch participants and interactions between them.
This model does not systematically incorporate pitch location, meaning that the resulting estimates of framing effects are confounded by the location just like \citet{Turkenkopf2008}'s.
To our knowledge, the most systematic attempt to study framing to date is \citet{JudgePavlidisBrooks2015}.
They introduce a mixed-effects probit regression model built in two stages: first, they estimate a baseline called strike probability using a proprietary model that accounts for location, count, handedness of batter, and ballpark. 
They then fit a probit regression model with a fixed effect for this baseline estimate and random effects for the pitch participants.  Underpinning their model is the curious assumption that the probit transformed called strike probability is \textit{linear} in the baseline probability estimate.
This assumption can over-leverage pitches with baseline probabilities close to 0 or 1 (e.g. pitches straight over home plate or several inches outside the strike zone) by arbitrarily inflating the associated intercept and slopes in the final probit model.
This can potentially result in highly unstable parameter estimates. 
%Curiously, they model the probit transformation of the called strike probability with an existing, proprietary estimate of the same probability, in essence linearizing the probit function. 
%This modeling choice lacks statistical justification and potential consequences on the validity of and uncertainty in the final estimates of framing impact are unknown and unexplored. 
%Further, it is not clear whether this existing estimate was computed with the same data used to fit the final probit model, further complicating the uncertainty quantification for their final framing estimates. 
Both \citet{JudgePavlidisBrooks2015}'s and \citet{Marchi2011}'s models unrealistically assume umpires differ only in some base-rate of calling strikes and that effect of factors like pitch location, count influence, and players is constant across umpires.
In light of this, we will proceed by fitting a separate model for each umpire.

Before proceeding, we introduce some notation. 
For a given taken pitch, let $y = 1$ if it is called a strike and let $y = 0$ if it is called a ball. 
Let $\mathbf{b}, \mathbf{ca}, \mathbf{co}, \mathbf{p}$ and $\mathbf{u}$ be indices corresponding to the batter, catcher, count, pitcher, and umpire for that pitch. 
Further, let $x$ and $z$ be the horizontal and vertical coordinates of the pitch as it crosses the front plane of home plate, respectively. 
To accommodate a separate model for each umpire $u$ we introduce vectors $\Theta^{u,B}, \Theta^{u,CA}, \Theta^{u,P},$ and $\Theta^{u,CO}$ to hold the \textit{partial effect} of each batter, catcher, count, and pitcher, respectively, on umpire $u$'s likelihood to call a strike. 
For each umpire $u,$ we introduce a function of pitch location, $f^{u}(x,z)$, that we will specify in more detail later.
At a high level, we model
\begin{equation}
\log{\left(\frac{\P(y = 1)}{\P(y = 0)}\right)} = \Theta^{\mathbf{u},B}_{\mathbf{b}} + \Theta^{\mathbf{u},CA}_{\mathbf{ca}} + \Theta^{\mathbf{u},P}_{\mathbf{p}} + \Theta^{\mathbf{u},CO}_{\mathbf{co}} + f^{\mathbf{u}}(x,z)
\label{eq:general_model}
\end{equation}

We leverage high-resolution pitch tracking data from the PITCHf/x system, described briefly in Section~\ref{sec:pitchfx}, to estimate how much a catcher influences umpires' chances of calling strikes and how large an impact such effects have on his team's success.
In Section~\ref{sec:models}, we introduce several simplifications of the model in Equation~\ref{eq:general_model} that still elicit umpire-to-umpire heterogeneity.
All of these models are fit in a hierarchical Bayesian framework, which provides natural uncertainty quantification for our framing estimates.
Such quantification, notably absent in previous framing studies, is vital, considering the fact that several teams are making framing-based roster decisions \citep{Drellich2014, Holt2014, Sanchez2015}.
We compare the predictive performances of these models in Section~\ref{sec:model_comparison} and assess the extent to which incorporating umpire-specific count and player effects lead to overfitting. 
We then translate our estimates of catcher effects from the log-odds scale to the more conventional scale of average runs saved.
We introduce two metrics in Section~\ref{sec:framing_impact} to estimate the impact framing has on team success.
We conclude with a discussion and outline several potential extensions of our modeling efforts.

\section{Data and Model}
\label{sec:data_model}

We begin this section with a brief overview, adapted primarily from \citet{Fast2010} and \citet{SidhuCafo2014}, of our pitch tracking dataset before introducing the hierarchical logistic regression model used to estimate each umpire's called strike probability. 

\subsection{PITCHf/x Data}
\label{sec:pitchfx}

%% Refine this sentence substantially

In 2006, the TV broadcast company Sportvision began offering the PITCHf/x service to track and digitally record the full trajectory of each pitch thrown using a system of cameras installed in major league ballparks.  
During the flight of each pitch, these cameras take 27 images of the baseball and the PITCHf/x software fits a quadratic polynomial to the 27 locations to estimate its trajectory \citep{SidhuCafo2014}.
This data is transmitted to the MLB Gameday application, which allows fans to follow the game online \citep{Fast2010}.
In addition to collecting pitch trajectory data, an MLB Advanced Media employee records game-state information during each pitch.
For instance, he or she records the pitch participants (batter, catcher, pitcher, and umpire) as well as the outcome of the pitch (e.g, ball, swinging strike, hit), the outcome of the at-bat (e.g. strikeout, single, home run), and any other game action (e.g. substitutions, baserunners stealing bases).
The PITCHf/x system also reports the approximate vertical boundaries of the strike zone for each pitch thrown. 
Taken together, the pitch tracking data and game-state data provide a high-resolution pitch-by-pitch summary of the game, available through the MLB Gameday API.
%This data is available through \href{http://gd2.mlb.com/components/game/mlb/}{MLB Gameday API}.

Though our main interest in this paper is to study framing effects in the 2014 season, we collected all PITCHf/x data from the 2011 to 2015 regular season.
In Section~\ref{sec:pitch_location}, we use the data from the 2011 -- 2013 seasons to select the function of pitch location $f^{u}(x,z)$ from Equation~\ref{eq:general_model}.
We then fit our model using the 2014 data and in Section~\ref{sec:model_comparison}, we assess our model's predictive performance using data from 2015.
In the 2014 season, there were a total of 701,490 pitches, of which 355,293 (50.65\%) were \textit{taken} (i.e. not swung at) and of these taken pitches, 124,642 (35.08\%) were called strikes.
Rather than work with all of the taken pitches, we restrict our attention to those pitches that are ``close enough'' to home plate to be ``frameable.''
More precisely, we first approximate a crude ``average rule book strike zone'' by averaging the vertical strike zone boundaries recorded by the PITCHf/x system across all players and all pitches, and then focused on the $N = 308,388$ taken pitches which were within one foot of this approximate strike zone. 
In all, there were a total of $n_{U} = 93$ umpires, $n_{B} = 1010$ batters, $n_{C} = 101$ catchers, and $n_{P} = 719$ pitchers.

\subsection{Adjusting for Pitch Location}
\label{sec:pitch_location}

Intuitively, pitch location is the main driver of called strike probability.
% and any credible attempt to estimate called strike probability must account for pitch location in some systematic way.
The simplest way to incorporate pitch location into our model would be to include the horizontal and vertical coordinates $(x,z)$ recorded by the PITCHf/x system as linear predictors so that $f^{u}(x,z) = \theta^{u}_{x}x + \theta^{u}_{z}z,$ where $\theta^{u}_{x}$ and $\theta^{u}_{z}$ are parameters to be estimated. 
While simple, this forces an unrealistic left-to-right and top-to-bottom monotonicity in the called strike probability surface. 
Another simple approach would be to use a polar coordinate representation, with the origin taken to be the center of the approximate rule book strike zone.
While this avoids any horizontal or vertical monotonicity, it assumes that, all else being equal, the probability of a called strike is symmetric around this origin. 

\begin{figure}[!h]
\centering
\includegraphics{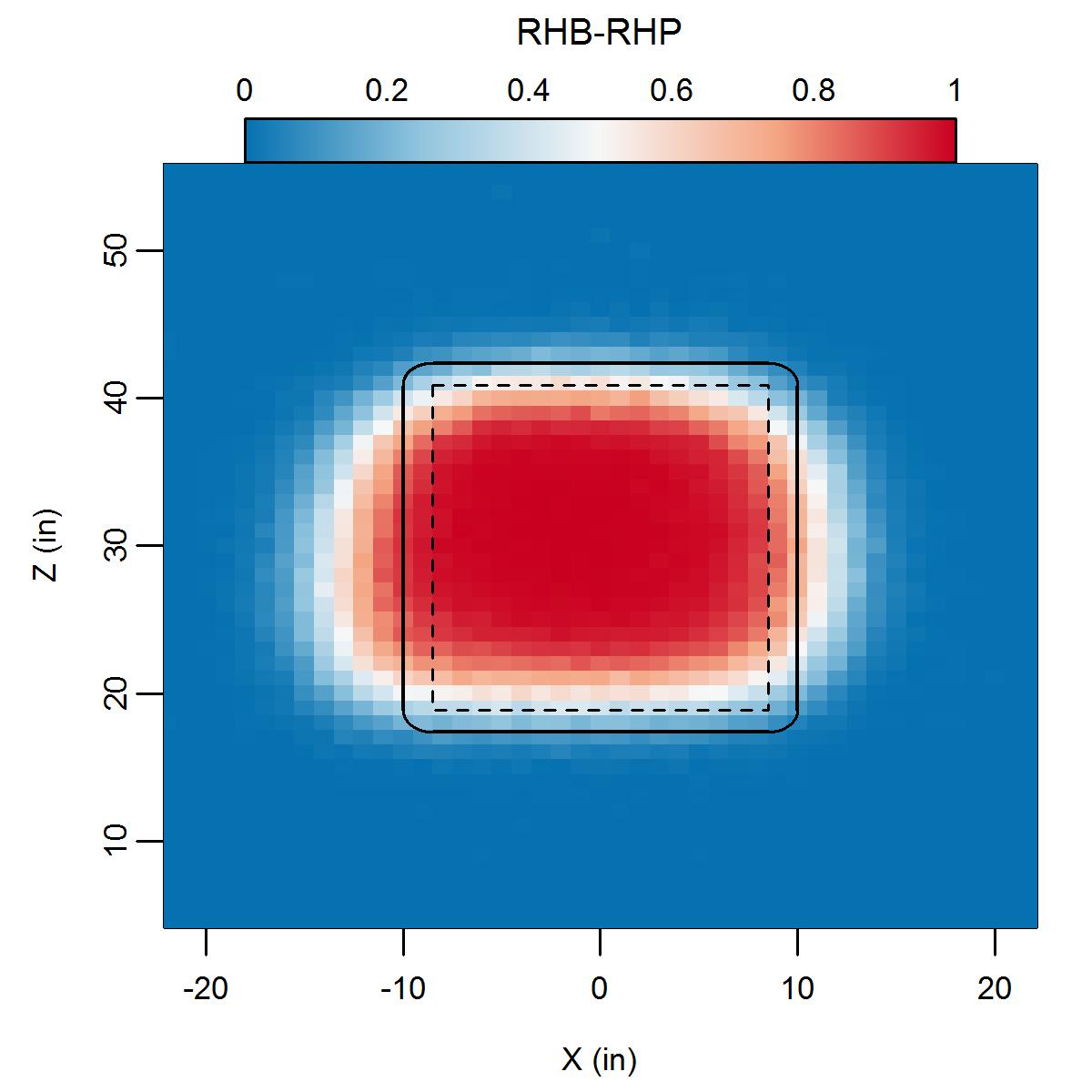}
\caption{Heat map of empirical called strike probabilities, aggregate over the three-year window 2011 -- 2013. The boundary of the approximate 2014 rule book strike zone is shown in dashed line. If the center of the pitch passes through the region bounded by the solid line, some part of the pitch passes through the approximate strike zone. Red = 100\% called strike probability, white = 50\%, and blue = 0\%.}
\label{fig:heatMap_2011_2013}
\end{figure}

Such symmetry is not observed empirically, as seen in Figure~\ref{fig:heatMap_2011_2013}, which divides the plane above home plate into 1" squares whose color corresponds to the proportion of pitches thrown in the three year window 2011 -- 2013 that pass through the square that are called strikes. 
Also shown in Figure~\ref{fig:heatMap_2011_2013} is the average rule book strike zone, demarcated with the dashed line, whose vertical boundaries are the average of the top and bottom boundaries recorded by the PITCHf/x system.
If the center of the pitch passes through the region bound by the solid line, then some part of the pitch passes through the approximate strike zone. 
This heat map is drawn from the umpire's perspective so right handed batters stand to the left of home plate (i.e. negative X values) and left-handed batters stand to the right (i.e. positive X values).
We note that the bottom edge of the figure stops 6 inches off of the ground and the left and right edges end 12 inches away from the edges of home plate.
Typically, batters stand an additional 12 inches to the right or left of the region displayed. 
Interestingly, we see that the empirical called strike probability changes rapidly from close to 1 to close to 0 in the span of only a few inches.

Rather than specifying a explicit parametrization in terms of the horizontal and vertical coordinates, we propose using a smoothed estimate of the historical log-odds of a called strike as an implicit parametrization of pitch location.
This is very similar to the model of \citet{JudgePavlidisBrooks2015}, who included the estimated called strike probability as a covariate in their probit model.

\begin{figure}[!h]
\centering
\includegraphics{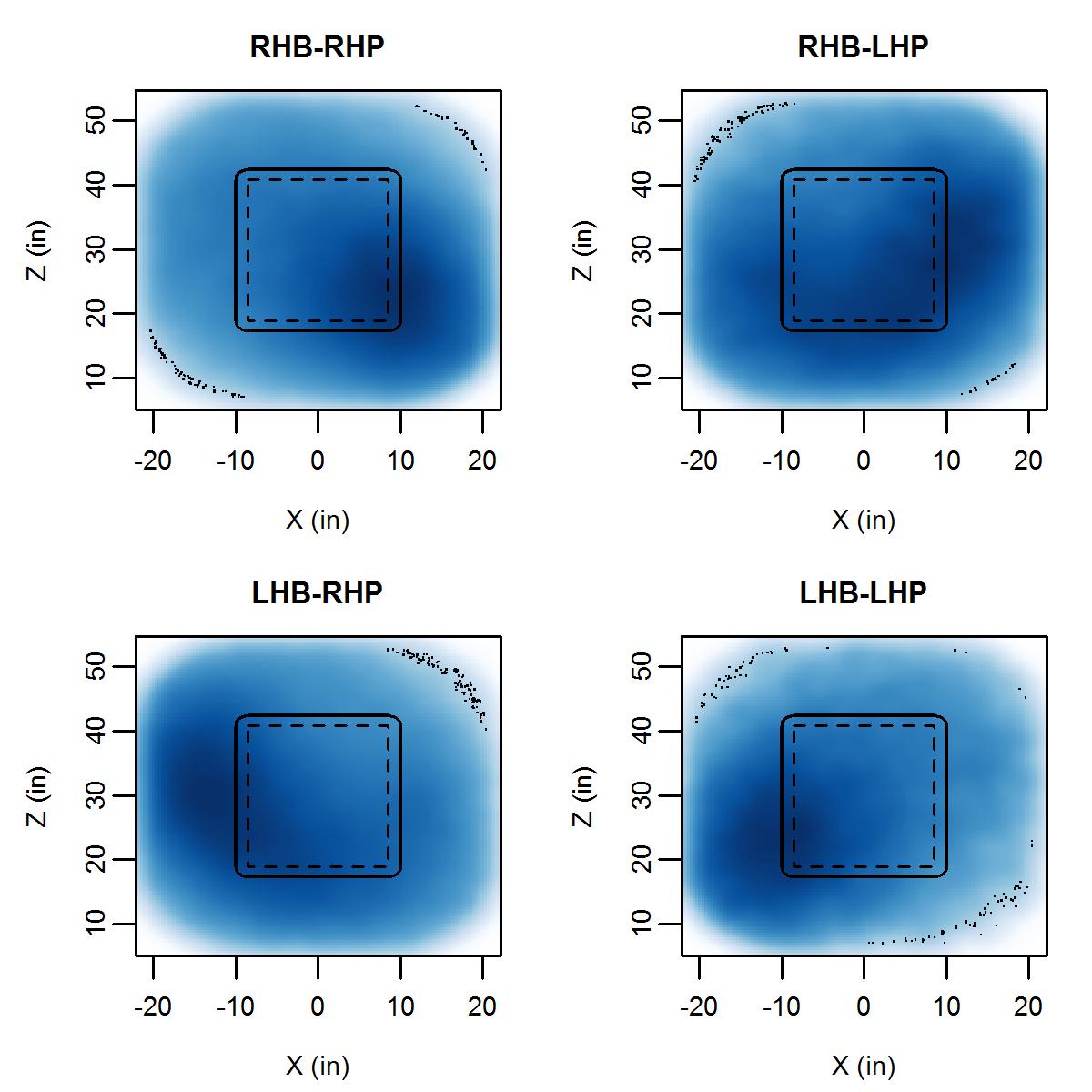}
\caption{Kernel density estimate of pitch location based on batter and pitcher handedness. Figures drawn from the umpires perspective so right-handed batters stand to the left of the displayed strike zone. Darker regions correspond to a higher density of pitches thrown to those locations.}
\label{fig:pitch_location_handedness}
\end{figure}

Figure~\ref{fig:pitch_location_handedness} plots the spatial distribution of taken pitches broken down by the batter and pitcher handedness.
Once again, the plots are drawn from the umpires' perspective so that a right handed batter stands to the left side of the figure and vice versa.
We see immediately that the spatial distribution of taken pitches varies considerably with the combination of batter and pitcher handedness.
When the batter and pitcher are of the same handedness, we see a decidedly higher density of ``low and outside'' pitches near the bottom corner of the average rule book strike furthest away from the batter.
In contrast, in the matchup between left-handed batters and right-handed pitchers, we see a higher density of pitches thrown to the outside edge of the strike zone further away from the batter.
%Interestingly, for the right-handed batters -- left-handed pitchers matchup, in addition to a high density of outside pitches, we see a considerably more inside pitches as well.
The differences in spatial distribution of pitches seen in Figure~\ref{fig:pitch_location_handedness} motivate us to use a separate smoothed estimate of the historical log-odds of a called strike for each combination of batter and pitcher handedness.

\begin{figure}[!h]
\centering
\includegraphics{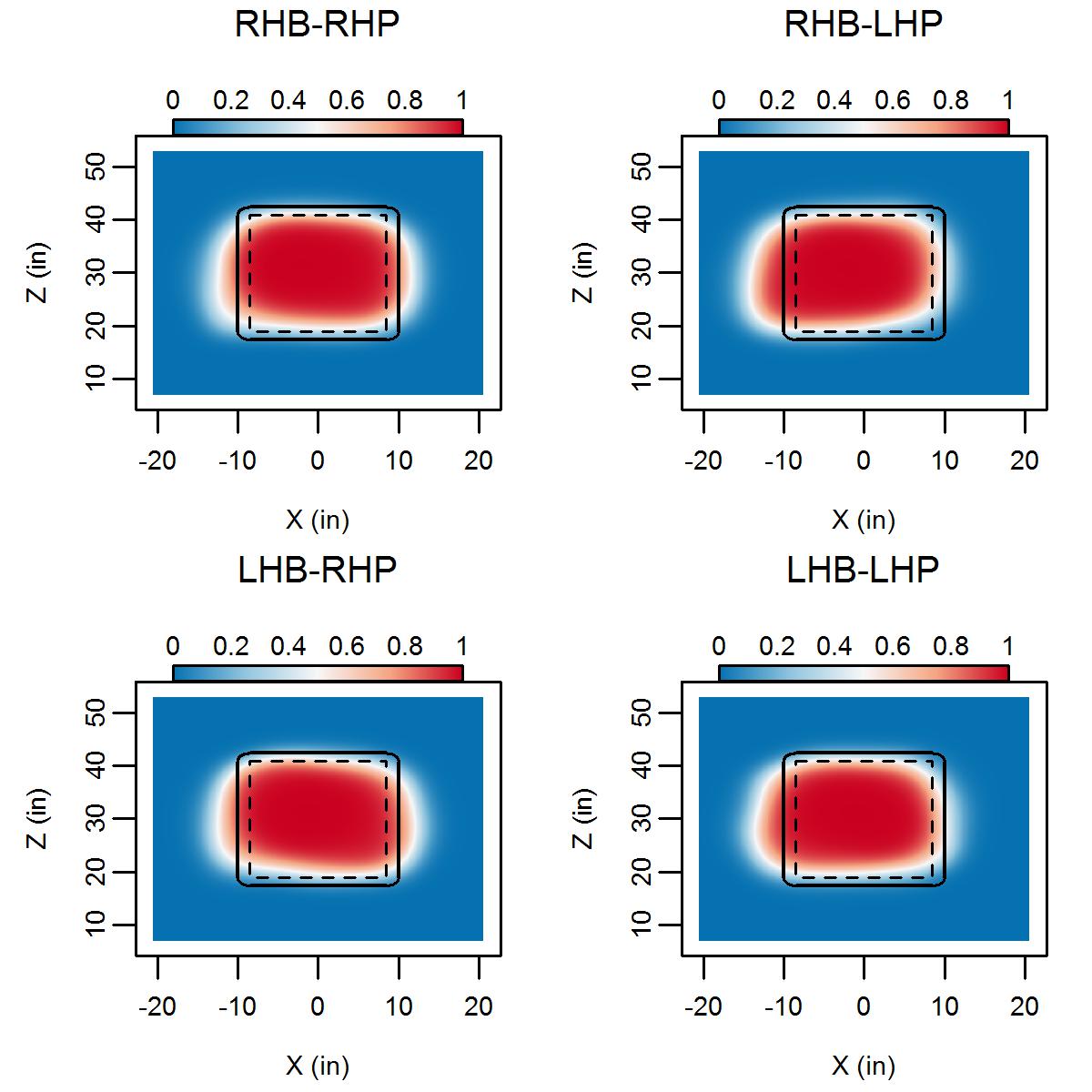}
\caption{hGAM forecasts based on batter and pitcher handedness. Red = 100\% called strike probability, white = 50\%, and blue = 0\%.}
\label{fig:hgam_sz_plots}
\end{figure}

Inspired by \citet{Mills2014}, we fit generalized additive models with a logistic link to the data aggregated from 2011 -- 2013, one for each combination of pitcher and batter handedness, hereafter referred to as the ``hGAMs'' or ``historical GAMS.'' 
These models express the log-odds of a called strike as a smooth function of the pitch location. 
Figures~\ref{fig:hgam_sz_plots} shows the hGAM forecasted called strike probabilities.
Interestingly, we see that for right handed pitchers, the corresponding hGAMs called strike probability surfaces very nearly align with the average rule book strike zone.
For left-handed pitchers, however, the hGAMs forecast a high called strike probability several inches to the left of the average rule book strike zone. 
This is perhaps most prominent for the matchup between right-handed batters and left-handed pitchers.
For each taken pitch in 2014 dataset, we used the appropriate hGAM to estimate the historical log-odds that the pitch was called a strike. 
We then use these estimates as continuous predictors in our model, so that potential player effects and count effects may be viewed as adjustments to these historical baselines.

%As seen in Figure~\ref{fig:pitch_location_handedness}, however, pitch location can vary considerably based on the handedness of the batter and pitcher.
%Interestingly, in matchups but the one between a right-handed batter and a left-handed pitcher, there is a high density of pitches thrown to the 
%Because of this, we actually fit four separate GAM's, hereafter referred to as the ``hGAMs'' or ``historical GAMs'', that express the log-odds of a called strike as a smooth function of the pitch location, one each for the four combinations of pitcher and batter handedness.
%Figure~\ref{fig:hgam_sz_plots} shows the four estimated called strike probability surfaces. 
%Recall that these figures are drawn from 

%We see that between 2011 and 2013, there are clear differences in how umpires called strikes, depending on combination of pitcher and batter handedness.

\subsection{Bayesian Logistic Regression Models}
\label{sec:models}

%As alluded to in Section~\ref{sec:introduction}, we consider several models of increasing complexity to predict umpires' calls in 2014. 
Before fully specifying our models, we label the 93 umpires $u_{1}, \ldots, u_{93}.$
Consider the $i^{th}$ called pitch and let $y_{i} = 1$ if it is called a strike and $y_{i} = 0$ if it is called a ball.
Let $h_{i}$ be a vector of length four, encoding the combination of batter and pitcher handedness on this pitch and let $\textbf{LO}_{i}$ be a vector of length four, containing three zeros and the estimated log-odds of a strike from the appropriate historic GAM based on the batter and pitcher handedness. 
Letting $x_{i}$ and $z_{i}$ denote the PITCHf/x coordinates of this pitch, we take $f^{u}(x_{i}, z_{i}) = h_{i}^{\top}\Theta^{u}_{0} + \textbf{LO}_{i}^{\top}\Theta^{u}_{LO},$ where $\Theta^{u}_{LO}$ is a vector of length four recording the partial effect of location and $\Theta^{u}_{0}$ is a vector of length four containing an intercept term, one for each combination of batter and pitcher handedness. 
Finally, let $u(i)$ denote which umpire called this pitch. 
To place all of the variables in our model on roughly similar scales, we first re-scale the corresponding historical GAM estimates for each combination of batter and pitcher handedness to have standard deviation 1. 

%Further let $h^{u}_{i}$ be a vector of length four, encoding the combination of batter and pitcher handedness on this pitch and let $\text{lo}_{i}$ be a vector of length four
%Further, let $h^{u}_{i} \in \left\{RR, RL, LR, LL \right\}$ be the combination of batter and pitcher handedness on this pitch and let $\text{lo}_{i}$ be the fitted log-odds of a called strike from the appropriate historical GAM. 
Finally let $\textbf{CO}_{i}, \textbf{CA}_{i}, \textbf{P}_{i}$ and $\textbf{B}_{i}$ be vectors encoding the count, catcher, pitcher, and batter involved with this pitch, and let $\Theta^{u}_{CO}, \Theta^{u}_{CA}, \Theta^{u}_{P},$ and $\Theta^{u}_{B}$ be vectors containing the partial effect of count, catcher, pitcher, and batter on umpire $u.$
For identifiability, we specify a single catcher, Brayan Pena, and count, 0 -- 0, as baseline values.
We can re-write the model from Equation~\ref{eq:general_model} as
$$
\log{\left(\frac{\P(y_{i} = 1)}{\P(y_{i} = 0)}\right)} = h_{i}^{\top}\Theta^{u(i)}_{0} + \textbf{LO}_{i}^{\top}\Theta^{u(i)}_{LO} + \textbf{CO}_{i}^{\top}\Theta^{u(i)}_{CO} + \textbf{CA}_{i}^{\top}\Theta^{u(i)}_{CA} + \textbf{P}_{i}^{\top}\Theta^{u(i)}_{P} + \textbf{B}_{i}^{\top}\Theta^{u(i)}_{B}
$$

%For identifiability, we designate 0--0 as a baseline count, Brayan Pena as the baseline catcher, Mike Leake as the baseline pitcher, and Melvin Upton as the baseline batter. 
%For identifiability, we specify a single catcher, Brayan Pena, and count, 0--0, as baselines values.
%In this way, the vector of intercepts, $\Theta^{u}_{0}$, capture the log-odds of umpires $u$ calling a strike on an 0--0 pitch caught by Pena  and thrown at a location where umpires historically called balls and strikes at equal rates. 

We are now ready to present several simplifications of this general model of gradually increasing complexity.
We begin first by assuming that the players and count have no effect on the log-odds of a called strike (i.e. that $\Theta^{u}_{CO}, \Theta^{u}_{CA}, \Theta^{u}_{P},$ and $\Theta^{u}_{B}$ are all equal to the zero vector for each umpire).
This model, hereafter referred to as Model 1, assumes that the only relevant predictor of an umpire's ball/strike decision is the pitch location but allows for umpire-to-umpire heterogeneity. 
We model, \textit{a priori},
\begin{eqnarray*}
\Theta^{u_{1}}_{0}, \ldots, \Theta^{u_{93}}_{0} | \Theta_{0} \sim N\left(\Theta_{0}, \tau^{2}_{0}I_{4}\right) \\  \Theta^{u_{1}}_{LO}, \ldots, \Theta^{u_{93}}_{LO} | \Theta_{LO} \sim N\left(\Theta_{LO}, \tau^{2}_{LO}I_{4}\right)\\
\Theta_{0} | \sigma_{0}^{2} \sim N\left(0_{4}, \sigma_{0}^{2}I_{4}\right)  \\
\Theta_{LO} | \sigma^{2}_{LO} \sim N\left(\mu_{LO}, \sigma^{2}_{LO}I_{4}\right)
\end{eqnarray*}

The vector $\mu_{LO}$ is taken to be the vector of standard deviations of the hGAM forecast for each combination of batter and pitcher handedness.
In this way, Model 1 centers the prior distribution of the log-odds of a strike at the hGAM forecasted log-odds. 
We may interpret the parameters $\tau^{2}_{0}$ and $\tau^{2}_{LO}$ as capturing the umpire-to-umpire variability in the intercept and location effects and we may view $\Theta_{0}$ and $\Theta_{LO}$ as the mean intercept and location effects averaged over all umpires.
By placing a further level of prior hierarchy on $\Theta_{0}$ and $\Theta_{LO},$ we render the $\Theta_{0}^{u}$'s and $\Theta_{LO}^{u}$'s dependent, both \textit{a priori} and \textit{a posteriori}.
In this way, while we are fitting a separate model for each umpire, these models are ``mutually informative'' in the sense that the estimate of umpire $u$'s intercept vector $\Theta^{u}_{0}$ will, for instance, be ``shrunk'' towards the average of all umpires' intercept vectors by an amount controlled by $\tau^{2}_{0}$ and $\tau^{2}_{LO}.$
Further priors on the hyper-parameters $\sigma^{2}_{0}$ and $\sigma^{2}_{LO}$ introduce dependence between the
components of $\Theta^{u}_{0}$ and $\Theta^{u}_{LO}$ as well, enabling us to ``borrow strength'' between the four combination of batter and pitcher handedness. 
%In this way, our posterior estimate of umpire $u$'s intercept for the right handed batter/right handed pitcher combination will, in addition to the observed data, depend on our estimate for the same umpire's intercept for the other three handedness combinations, and our estimates of every other umpires' intercepts for all handedness combinations to a somewhat lesser extent.

While Model 1 essentially estimates a separate called strike probability surface for each umpire, it entirely precludes the possibility of player or count effects. 
We now consider two successive expansions of Model 1.
In Model 2, we incorporate both catcher and count effects that are assumed to be constant across umpires.
That is, in Model 2 we assume that all of the the $\Theta^{u}_{CO}$'s are all equal to some common value $\Theta_{CO}$ and all of the $\Theta^{u}_{CA}$'s are equal to some common value $\Theta_{CA}.$
Similarly, in Model 3 we augment Model 2 with constant pitcher effects and constant batter effects. 
\textit{A priori}, we model 
$$
\Theta_{CO} | \sigma^{2}_{CO} \sim N\left(0_{11}, \sigma^{2}_{CO}I_{11}\right)
$$
and consider similar, zero-mean spherically-symmetric Gaussian priors for $\Theta_{CA}, \Theta_{P}$ and $\Theta_{B},$ while retaining the same prior specification on the $\Theta^{u}_{0}$'s and $\Theta^{u}_{LO}$'s.
Though they elaborate on Model 1, Models 2 and 3 still represent a vast simplification to the general model in Equation~\ref{eq:general_model} as they assume that there is no umpire-to-umpire variability in the count or player effects.
This leads us to consider Model 4, which builds on Model 2 by allowing umpire-specific count and catcher effects, and Model 5, which includes umpire-specific batter and pitcher effects and corresponds to the general model in Equation~\ref{eq:general_model}
We model
\begin{eqnarray*}
\Theta^{u_{1}}_{CO}, \cdots, \Theta^{u_{93}}_{CO} | \Theta_{CO},\tau_{CO}^{2} \sim N\left(\Theta_{CO}, \tau^{2}_{CO}I_{11}\right) \\
\Theta^{u}_{CO} | \sigma^{2}_{CO} \sim N\left(0_{11}, \sigma^{2}_{CO}I_{11}\right)
\end{eqnarray*}
and consider similarly structured prior hierarchies for $\Theta^{u}_{CA}, \Theta^{u}_{B}, \Theta^{u}_{P}$ in Models 4 and 5.
Throughout, we place independent Inverse Gamma(3,3) hyper-priors on the top-level variance parameters $\sigma^{2}_{0}, \sigma^{2}_{LO}, \sigma^{2}_{CO}, \sigma^{2}_{CA}, \sigma^{2}_{P}$ and $\sigma^{2}_{B}.$

It remains to specify the hyper-parameters $\tau^{2}_{0}, \tau^{2}_{LO}, \tau^{2}_{CO}, \tau^{2}_{CA},\tau^{2}_{P}$ and $\tau^{2}_{B}$ which capture the umpire-to-umpire variability in the intercept, location, count, and player effects.
For simplicity, we fix these hyper-parameters to be equal to $0.25$ in the appropriate models.
To motivate this choice, consider how two umpires would call a pitch thrown at a location where the historical GAM forecasts a 50\% called strike probability.
According to Model 2, the difference in the two umpires' log-odds of a called strike follows a $N\left(0, 2(\tau_{0}^{2} + \tau^{2}_{CO} + \tau^{2}_{CA})\right)$ distribution, \textit{a priori}.
Taking $\tau_{0}^{2} = \tau^{2}_{CO} = \tau^{2}_{CA} = 0.25$ reflects a prior belief that there is less than a 10\% chance that one umpire would call a strike 75\% of the time while the other calls it a strike only 25\% of the time. 
For simplicity, we take $\tau^{2}_{LO}  = \tau^{2}_{B} = \tau^{2}_{P} = 0.25$ as well.

\section{Model Performance and Comparison}
\subsection{Predictive Performance}
\label{sec:model_comparison}

% Fully saturated model
% 179,011 seconds to sample .... about 50 hours.

We fit each model in Stan \citep{Stan} and ran two MCMC chains for each model.
All computations were done in R (versions 3.3.2 and later) and the MCMC simulation was carried out in RStan (versions 2.14.1 and later) on a high-performance computing cluster.
For each model, after burning-in the first 2000 iterations, the Gelman-Rubin $\hat{R}$ statistic for each parameter was less than 1.1, suggesting convergence.
We continued to run the chains, after this burn-in, until each parameter's effective sample size exceeded 1000.
For Models 1 and 2, we found that running the sampler for 4000 total iterations was sufficient while for Models 3, 4, and 5, we needed 6000 iterations.
The run time of these samplers ranged from just under an hour (Model 1) to 50 hours (Model 5). 

Using the simulated posterior draws from each model, we can approximate the mean of the posterior predictive distribution of the called strike probability for each pitch in our 2014 dataset. 
Table~\ref{tab:inSample_error} shows the misclassification and mean square error for Models 1 -- 5 over all pitches from 2014 and over two separate regions, as well as the error for the historical GAM forecasts. 
Region 1 consists of all pitches thrown within 1.45 inches on either side of the boundary of the average rule book strike zone defined in Section~\ref{sec:pitchfx}.
Since the radius of the ball is about 1.45 inches, the pitches in Region 1 are all ``borderline'' calls in the sense that only part of the ball passes through the strike zone but are, by rule, strikes.
Region 2 consists of all pitches thrown between 1.45 and 2.9 inches outside the boundary of the average rule book strike zone. 
These pitches miss the strike zone by an amount between one and two ball's width, and ought to be called balls by the umpire.
To compute misclassification error, we used 0.5 as the threshold for a strike.

\begin{table}[!h]
\centering
\caption{In-sample predictive performance for several models}
\label{tab:inSample_error}
\footnotesize
\begin{tabular}{llcccccc} \\\hline
~  & ~ &Model 1 &Model2 & Model 3 & Model 4 & Model 5 & hGAM \\ 
~ & $\#$ Parameters & 744 & 855 & 2582 & 11,067 & 171,168 & -- -- \\ \hline
Overall & MISS & 0.103 & 0.100 & 0.099 & 0.096 & \bf 0.0856 & 0.105 \\
~ & MSE & 0.073 & 0.071 & 0.069 & 0.068 & \bf 0.061 & 0.074 \\
Region 1 & MISS & 0.248 & 0.236 & 0.232 & 0.225 & \bf 0.195 & 0.258 \\
~ & MSE & 0.163 & 0.156 & 0.153 & 0.150 & \bf 0.133 & 0.168 \\
Region 2 & MISS & 0.214 & 0.209 & 0.205 & 0.203 & \bf 0.184 & 0.215 \\
~ & MSE & 0.153 & 0.149 & 0.146 & 0.144 & \bf 0.129 & 0.156 \\ \hline
\end{tabular}
\end{table}

We see that Models 1 -- 5 outperform the historical GAMs overall and in both Regions 1 and 2.
This is hardly surprising, given that the hGAMs were trained on data from 2011 -- 2013 and the other models were trained on the 2014 data.
Recall that Model 1 only accounted for pitch location.
As we successively incorporating count and catcher (Model 2), and pitcher and batter (Model 3) effects, we find that the overall error drops.
Finally, Model 5 has the best performance across the board.
This is entirely expected as Model 5 given the tremendous number of parameters. 

Of course, we would be remiss if we assessed predictive performance only with training data.
Table~\ref{tab:outSample_error} compares such out-of-sample predictive performance, by considering pitches from the 2015 season for which the associated batter, catcher, pitcher, and umpire all appeared in our 2014 dataset.

\begin{table}[!h]
\centering
\footnotesize
\caption{Out-of-sample predictive performance for several models}
\label{tab:outSample_error}

\begin{tabular}{llcccccc} \\\hline
~  & ~ &Model 1 &Model2 & Model 3 & Model 4 & Model 5 & hGAM \\ 
~ & $\#$ Parameters & 744 & 855 & 2582 & 11,067 & 171,168 & -- -- \\ \hline
Overall & MISS & 0.107 & 0.105 & \bf 0.105 & 0.106 & 0.106 & 0.109 \\
~ & MSE & 0.075 & 0.074 & \bf0.074 & 0.075 & 0.074 & 0.076 \\
Region 1 & MISS & 0.256 & 0.245 & \bf0.244 & 0.248 & 0.246 & 0.267 \\
~ &MSE & 0.167 & 0.162 & \bf0.161 & 0.163 & 0.162 & 0.173 \\
Region 2 & MISS & 0.236 & 0.232 & \bf0.231 & 0.233 & 0.234 & 0.237 \\
~ &MSE & 0.169 & 0.166 & \bf0.165 & 0.166 & 0.165 & 0.170 \\ \hline
\end{tabular}
\end{table}

Now we see that Model 3 has the best out-of-sample performance overall and in Regions 1 and 2. 
The fact that Models 4 and 5 have worse out-of-sample performance, despite having very good in-sample performance is a clear indication that these two over-parametrized models have overfit the data.
One could argue, however, that comparing predictive performance on 2015 data is not the best means of diagnosing overfitting.
\citet{Roegelle2014}, \citet{Mills2017a}, and \citet{Mills2017b} have documented year-to-year changes in umpires' strike zone enforcement ever since Major League Baseball began reviewing and grading umpires' decisions in 2009.
In Appendix~\ref{app:holdout}, we report the results from a cross-validation study, in which we repeatedly re-fit Models 1 -- 5 using 90\% of the 2014 data and assessing performance on the remaining 10\%, that similarly demonstrates Model 3's superiority.

Model 3's superiority over Models 1 and 5 reveals that although accounting for player effects can lead to improved predictions of called strike probabilities, we cannot reliably estimate an individual catchers catcher effects on individual umpires with a single season's worth of data.

%Table~\ref{tab:holdout_error} shows the average misclassification and mean square error over 10 such holdout sets.
%\begin{table}[!h]
%\centering

%\caption{Hold-out misclassification rate (MISS) and mean square error (MSE) for several models ($\times 100$)}
%\label{tab:holdout_error}
%\footnotesize
%\begin{tabular}{llccccc} \\ \hline
%~ & ~ Model 1 & Model 2 & Model 3 & Model 4 & Model 5 \\ 
%~ & $\#$ Parameters & 744 & 855 & 2582 & 11,067 & 171,168  \\ \hline
%Overall & MISS & 0.104 & 0.101 & \bf0.101 & 0.107 & 0.101 \\
%~ & MSE & 0.073 & 0.071 & \bf 0.071 & 0.072 & 0.072 \\
%Region 1 & MISS & 0.251 & \bf 0.239 & 0.240 & 0.242 & 0.240 \\
%~ & MSE & 0.164 & 0.158 & \bf 0.157 & 0.159 & 0.158 \\
%Region 2 & MISS & 0.213 & 0.208 & \bf0.206 & 0.208 & 0.207 \\
%~ & MSE & 0.154 & 0.150 & \bf0.148 & 0.150 & 0.149 \\ \hline
%\end{tabular}
%\end{table}
%Together, the results of Tables~\ref{tab:outSample_error} and~\ref{tab:holdout_error} clearly indicate that Models 4 and 5 overfit the data. 
%In the context of framing, overfitting called strike probabilities in a single year could skew retrospective estimates of the impact framing has.
%More importantly, it diminishes our ability to forecast the impact a catcher's framing has into the future, a key consideration for teams making roster decisions. 

\subsection{Full Posterior Analysis}
\label{sec:full_posterior_analysis}

\begin{figure}[!h]
\centering
\includegraphics{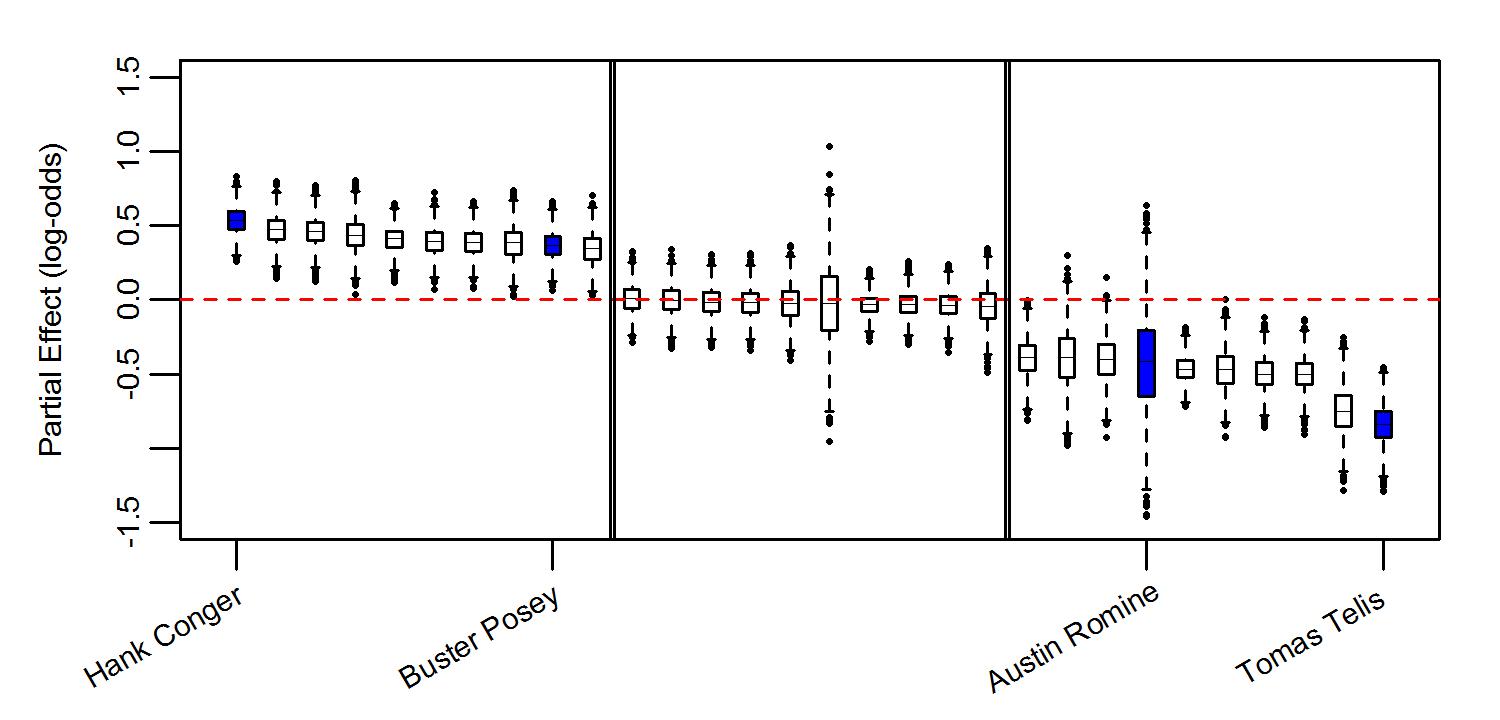}
\caption{Comparative box plots of 30 catcher effects sorted by the posterior mean of their partial effect on the log-odds scale}
\label{fig:catcher_boxplots}
\end{figure}

We now examine the posterior samples from Model 3 more carefully. 
%Figure~\ref{fig:catcher_boxplots} shows box plots of the posterior distributions of the catcher effects.
Figure~\ref{fig:catcher_boxplots} shows box plots of the posterior distributions of catcher effects on the log-odds scale for the catchers with the top 10 posterior means, the bottom 10 posterior means, and the middle 10 posterior means. 

We see that there are some catchers, like Hank Conger and Buster Posey, whose posterior distributions are entirely supported on the positive axis, indicating that, all else being equal, umpires are more likely to call strikes when they are caught by these catchers as opposed to the baseline catcher, Brayan Pena.
On the other extreme, there are some catchers like Tomas Tellis with distinctly negative effects.
As we would expect, catchers who appeared very infrequently in our dataset have very wide posterior distributions.
For instance, Austin Romine caught only 61 called pitches and his partial effect has the largest posterior variance among all catchers. 
It is interesting to see that all of the catcher effects, on the log-odds scale, are contained in the interval [-1.5,1.5], despite the prior placing nearly 20\% of its probability outside this interval. 
The maximum difference on the log-odds scale between the partial effects of any two catchers is 3, with high posterior probability.
For context, a change of 3 in log-odds corresponds to a change in probability from 18.24\% to 81.76\%.
As it turns out, the posterior distribution of each count effect is also almost entirely supported in the interval [-1.5, 1.5], on the log-odds scale.
This would seem to suggest that catcher framing effects are comparable in magnitude to the effect of count.
We explore this possibility in much greater detail in Appendix~\ref{app:catcher_count_effects}. 

\begin{figure}[!h]
\centering
\includegraphics{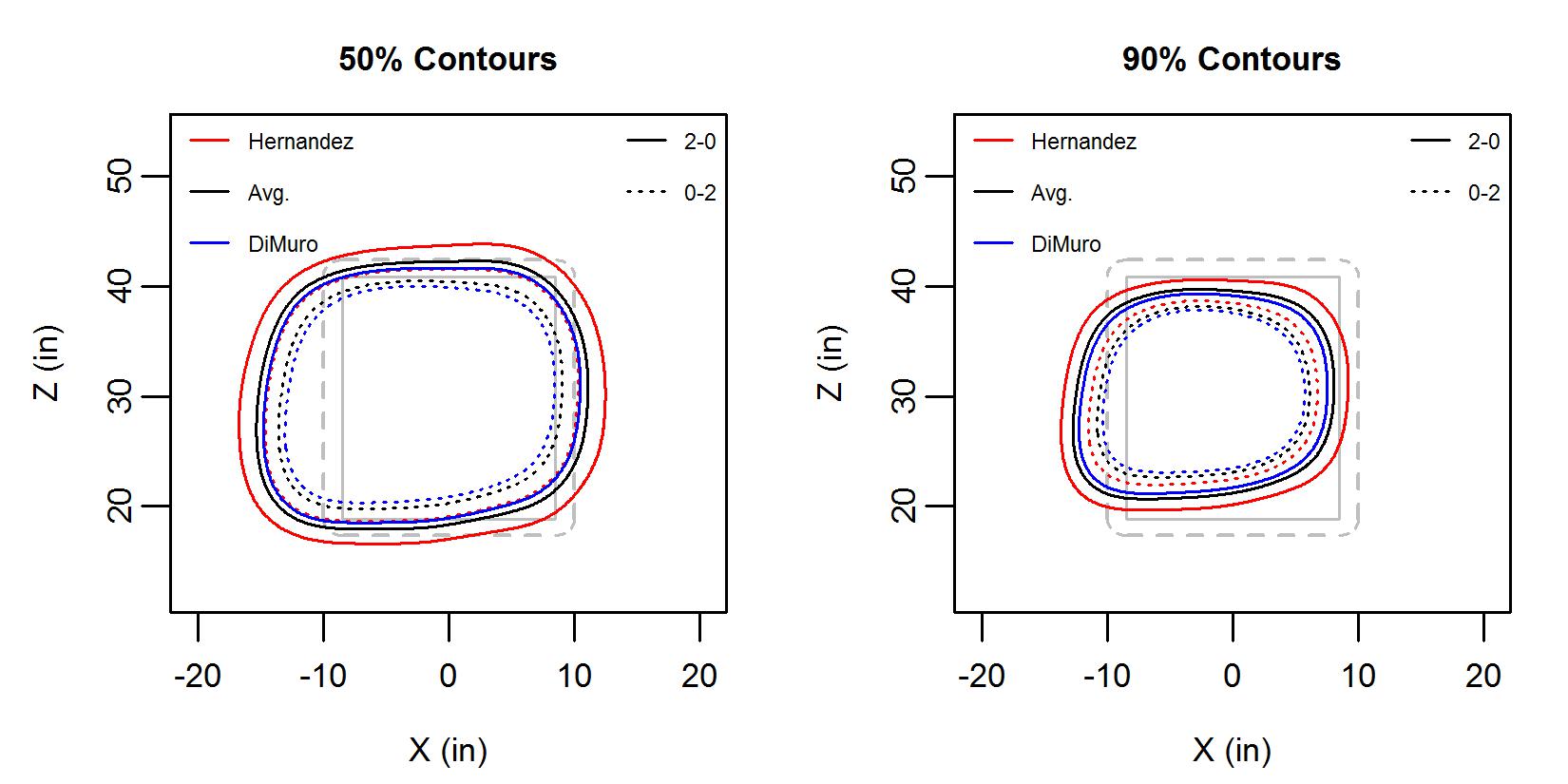}
\caption{50\% and 90\% contours for called strike probability for the Bumgarner-Puig-Posey matchup for different umpires in different counts.}
\label{fig:matchup_50_90}
\end{figure}

Armed with our simulated posterior draws, we can create posterior predictive strike zones for a given batter-pitcher-catcher-umpire matchup.
Suppose, for instance, that Madison Bumgarner is pitching to the batter Yasiel Puig, with Buster Posey catching. 
Figure~\ref{fig:matchup_50_90} shows the 50\% and 90\% contours of the posterior predictive called strike probability for two umpires, Angel Hernandez and Mike DiMuro, and an average umpire in a 2 -- 0 and 0 -- 2 count.
Note, if the center of the pitch passes within the region bounded by the dashed gray line in the figure, then some part of the ball passes through the average rule book strike zone, shown in gray.
Puig is a right-handed batter, meaning that he stands on the left-hand side of the approximate rule book strike zone, from the umpire's perspective.

Across the board, Hernandez's contours enclose more area than the average umpire's contours and DiMuro's contours enclose less area than the average umpire's.
For instance, on a 2 -- 0 count, Hernandez's 50\% contour covers 4.37 sq. ft., the average umpire's covers 3.87 sq. ft., and DiMuro's covers 3.53 sq. ft.
The contours on a 0 -- 2 pitch are much smaller, indicating that all else being equal, these umpires are less likely to call a strike on an 0 -- 2 pitch than on a 2 -- 0 pitch.
Each of the 50\% contours extend several inches to the left or \textit{inside} edge of the approximate rule book strike zone.
At the same time, the same contours do not extend nearly as far beyond the right or \textit{outside} edge of the strike zone.
This means that, Hernandez, DiMuro, and the average umpires are more likely to call strikes on pitches that miss the inside edge of the strike zone than they are on pitches that miss the outside edge by the same amount.
Even the 90\% contour on a 2 -- 0 count extend a few inches past the inside edge of the strike zone, implying that Hernandez, DiMuro, and the average umpire will almost always call a strike that misses the inside edge of a the strike zone so long as it is not too high or low. 
Interestingly, we see that the leftmost extents of DiMuro's and the average umpire's 90\% contours on a 0 -- 2 pitch nearly align with the dashed boundary on the inside edge.
A pitch thrown at this location will barely cross the average rule book strike zone, indicating that at least on the inside edge, DiMuro and umpires on average tend to follow the rule book prescription, calling strikes over 90\% of the time. 

The same is not true at the top, bottom, or outside edge.
For instance, the rightmost extent of average umpires' 90\% contour on a 0 -- 2 pitch lies several inches within the outside edge of the strike zone.
So in the space of about four and a half inches, the average umpires' called strike probability drops dramatically from 90\% to 50\%, despite the fact that according to the rule book these pitches should be called a strike.

Figure~\ref{fig:matchup_50_90} is largely consistent with the empirical observation that Hernandez tends to call a much more permissive strike zone than DiMuro: Hernandez called 42.67\% of taken pitches strikes (1624 strikes to 2182 balls) and DiMuro called 39.92\% of taken pitches strikes (1220 strikes to 1836 balls).
On 2 -- 0 pitches, Hernandez's strike rate increased to 51.44\% (71 strikes to 67 balls) and DiMuro's increased to 48.31\% (57 strikes to 67 balls). 

\section{Impact of framing}
\label{sec:framing_impact}

%We saw earlier that incorporating catcher effects can improve the fit, but allowing these catcher effects to vary by umpire tends to over-fit the data.
We now turn our attention now to measuring the impact framing has on the game.
Formally, let $S$ be a random variable counting the number of runs the pitching team gives up after the current pitch to the end of the half-inning.
Using slightly different notation than that in Section~\ref{sec:models}, let $\mathbf{h}$ encode the handedness of the batter and pitcher and let $\mathbf{lo}$ be  the estimated log-odds of a called strike from the appropriate historical GAM. 
Let $\mathbf{b}, \mathbf{ca}, \mathbf{co}, \mathbf{p}$ and $\mathbf{u}$ denote the batter, catcher, count, pitcher, and umpire involved in the pitch.
Finally, denote the baseline catcher Brayan Pena by $ca_{0}.$ 
For compactness, let $\xi = \left(\mathbf{u}, \mathbf{co}, \mathbf{lo}, \mathbf{b}, \mathbf{p}, \mathbf{h} \right)$ and observe that every pitch in our dataset can be identified by the combination $\left(\mathbf{ca}, \xi\right).$
For each catcher $ca,$ let $\mathcal{P}_{ca}$ be the set of all called pitches caught by catcher $ca$:
$$
\mathcal{P}_{ca} = \left\{ \left(\mathbf{ca}, \xi \right): \mathbf{ca} = ca \right\}.
$$
Finally, let $TAKEN$ be an indicator for the event that the current pitch was taken and let $CALL \in \left\{Ball, Strike\right\}$ be the umpire's ultimate call.
We will be interested in the expected value of $S,$ conditioned on $\left(\mathbf{ca}, \xi\right)$, the fact that pitch was taken, and the umpire's call.
Assuming, conditioned on the count, the fact that the pitch was taken, and the call, $S$ is independent of pitch location and participants, we have 

%\begin{equation}
%\label{eq:runs_given_up}
$$
\E[S |\mathbf{ca}, \xi, TAKEN] = \sum_{CALL}{\E[S | COUNT, TAKEN, CALL]\P\left(CALL|\mathbf{ca}, \xi, TAKEN\right)}
$$
%\end{equation}

To determine the expected number of runs given up that can be attributed to a catcher $ca$, we may consider the counter-factual scenario in which the catcher is replaced by the baseline catcher, Brayan Pena, with all other factors remaining the same.
In this scenario, the expected number of runs the fielding team gives up in the remaining of the half-inning is $E\left[S | \mathbf{ca} = ca_{0}, \xi, TAKEN, CALL\right].$
We may interpret the difference
$$
\E\left[S | \mathbf{ca} = ca, \xi, TAKEN, CALL\right] - \E\left[S | \mathbf{ca} = ca_{0}, \xi, TAKEN, CALL\right]
$$
as the average number of runs saved (i.e. negative of runs given up) by catcher $ca$'s framing, relative to the baseline. 
%how many fewer runs, on average, the fielding team gives up with the original catcher $c$ than it would with the baseline catcher.
%In other words, it is the average number of runs saved by catcher $c$'s framing, relative to the baseline.
A straightforward calculation shows that this difference is exactly equal to
%\begin{equation}
%\label{eq:runs_saved}
$$
f(ca, \xi) = \left(\P(Strike |\mathbf{ca} = ca, \xi, TAKEN) - \P(Strike| \mathbf{ca} = ca_{0}, \xi, TAKEN)\right) \times \rho(COUNT),
$$
%\end{equation} 
where
$$
\rho(COUNT) = \E[S|COUNT, TAKEN, Ball] - \E[S|COUNT, TAKEN, Strike].
$$

We can interpret the difference in called strike probabilities above as catcher $ca$'s \textit{framing effect}: it is precisely how much more the catcher adds to the umpires' called strike probability than the baseline catcher, over and above the other pitch participants, pitch location, and count.
We can easily simulate approximate draws from the posterior distribution of this difference using the posterior samples from Model 3.
We interpret $\rho$ as the value of a called strike in a given count: it measures how many more runs a team is expected to give up if a taken pitch is called a ball as opposed to a strike.

% Mention Albert(2010) here if needed.
To compute $\rho$,we begin by computing the difference in the average numbers of runs scored after a called ball and after a called strike in each count.
For instance, 182,405 0 -- 1 pitches were taken (140,667 balls, 41,738 called strikes) between 2011 and 2014. 
The fielding team gave up an average of 0.322 runs following a ball on a taken 0 -- 1 pitch, while they only gave up an average of 0.265 runs following a called strike on a taken 0 -- 1 pitch.
So conditional on an 0 -- 1 pitch being taken, a called strike saves the fielding team 0.057 runs, on average.
Table~\ref{tab:run_values} shows the number of runs scored after a called ball or a called strike for each count, as well as an estimate of $\rho.$
Also shown is the relative proportion of each count among our dataset of taken pitches from 2011 to 2014.
We see, for instance, that a called strike is most valuable on a 3 -- 2 pitch but only 2.1\% of the taken pitches in our dataset occurred in a 3 -- 2 count. 
This calculation is very similar to the seminal run expectancy calculation of \citet{Lindsey1963}, though ours is based solely on count rather than on the number of outs and base-runner configuration. 
\citet{Albert2010} also computes a count-based run expectancy, through his valuations are derived using the linear weights formula of \citet{ThornPalmer1985} rather than the simple average. 
See \citet{Albert2015} for a more in-depth discussion of run expectancy.

The weighted average run value of a called strike based on Table~\ref{tab:run_values} is 0.11 runs, slightly smaller than the value of 0.14 used by \citet{JudgePavlidisBrooks2015} and much smaller than the 0.161 figure used by \citet{Turkenkopf2008}.
The discrepancy stems from the fact that we estimated the run values based only on taken pitches while most other valuations of strikes include swinging strikes and strikes called off of foul balls.
It is worth stressing at this point that in our subsequent calculations of framing impact we use the count-based run valuation as opposed to the weighted average value.

\begin{table}[!h]
\centering
\footnotesize
\caption{Empirical estimates of run expectancy and run value, with standard errors in parentheses}
\label{tab:run_values}
\begin{tabular}{ccccc}
\hline
Count & Ball & Strike & Value of Called Strike $\rho$ & Proportion \\ \hline
0 -- 0 & 0.367 (0.002) & 0.305 (0.002) & 0.062 (0.002) &  36.2\%\\
0 -- 1 & 0.322 (0.002) & 0.265 (0.004)& 0.057 (0.004) & 12.5\% \\
0 -- 2 & 0.276 (0.003) & 0.178 (0.007) & 0.098 (0.008) & 5.5 \% \\
1 -- 0 & 0.427 (0.003) & 0.324 (0.003) & 0.103 (0.005) & 11.5\% \\
1 -- 1 & 0.364 (0.003) & 0.280 (0.004) & 0.084 (0.005) & 8.8\% \\
1 -- 2 & 0.302 (0.003) & 0.162 (0.006) & 0.140 (0.006) & 6.9\%\\
2 -- 0 & 0.571 (0.007) & 0.370 (0.006) & 0.201 (0.009) & 3.9\%\\
2 -- 1 & 0.468 (0.005) & 0.309 (0.006) & 0.159 (0.008) & 4.0\%\\
2 -- 2 & 0.383 (0.004) & 0.165 (0.006) & 0.218 (0.007) & 4.8\%\\
3 -- 0 & 0.786 (0.013) & 0.481 (0.008) & 0.305 (0.015) & 1.9\%\\
3 -- 1 & 0.730 (0.010) & 0.403 (0.009) & 0.327 (0.014) & 1.8\%\\
3 -- 2 & 0.706 (0.008) & 0.166 (0.008) & 0.540 (0.011) & 2.1\%\\ \hline
\end{tabular}
\end{table}

With our posterior samples and estimates of $\rho$ in hand, we can simulate draws from the posterior distribution of $f(ca,\xi)$ for each pitch in our dataset. 
An intuitive measure of the impact of catcher $ca$'s framing, which we denote RS for ``runs saved'' is
$$
RS(ca) = \sum_{(\mathbf{ca}, \xi) \in \mathcal{P}_{ca}}{f(ca, \xi)}.
$$

The calculation of RS is very similar to the one used by \cite{JudgePavlidisBrooks2015} to estimate the impact framing has on the game.
Rather than using fixed baseline catcher, \citet{JudgePavlidisBrooks2015} reports the difference in expected runs saved relative to a hypothetical average catcher.
According to their model, Brayan Pena, our baseline catcher, was no different than this average catcher, so our estimates of RS may be compared to the results of \citet{JudgePavlidisBrooks2015}.
Table~\ref{tab:runsSaved} shows the top and bottom 10 catchers, along with the number of pitches in our dataset received by the catchers, and the posterior mean, standard deviation, and 95\% credible interval of their RS values. 
Also shown are \citet{JudgePavlidisBrooks2015}'s estimates of runs saved for the catchers, as well as the number of pitches used in their analysis. 

\begin{table}[!h]
\centering
\footnotesize
\caption{Top and Bottom10 catchers according to the posterior mean of RS. The column BP contain \citet{JudgePavlidisBrooks2015}'s estimates that appeared on the Baseball Prospectus website}
\label{tab:runsSaved}
\begin{tabular}{llcccc} \hline
Rank & Catcher  & Runs Saved (SD)  & 95\% Interval     &  N       & BP\\ \hline
1 & Miguel Montero   & 25.71 (5.03) &[15.61, 35.09] &  8086  & 11.2 (8172) \\
2 & Mike Zunino        & 22.72 (5.17)  &[12.56, 32.31] &  7615 & 20.4 (7457) \\
3 & Jonathan Lucroy & 19.56 (5.69) &[8.16, 30.49]   & 8398 &16.4 (8241) \\
4 & Hank Conger      &19.34 (3.24) &[12.93, 25.65] & 4743 &  23.8 (4768) \\
5 & Rene Rivera       &18.81 (3.69)  & [11.63, 25.89] & 5091 & 22.5 (5182) \\
6 & Buster Posey      &17.01 (4.14) & [8.79, 25.01] & 6385 & 23.6 (6190) \\
7 & Russell Martin    & 14.35 (4.41) & [5.85, 22.77] &  6388 & 14.9 (6502) \\
8 & Brian McCann    &14.01 (3.95) &  [6.18, 21.66] & 6335 & 9.7 (6471) \\
9 & Yasmani Grandal & 12.88 (2.98) & [7.18, 18.69] & 4248 & 14.5 (4363)\\ 
10 & Jason Castro &12.61 (4.43) & [3.80, 21.08] & 7065 & 11.5 (7261) \\ \hline
92 & Josmil Pinto & -6.49 (1.41) & [-9.32, -3.76] & 1748 & -6.9 (1721)\\
93 & Welington Castillo & -6.70 (4.28) & [-15.19, 1.78] & 6667 &  -15.6 (6661) \\
94 & Chris Iannetta & -7.50 (4.46) & [-16.18, 1.08] & 6493 & -7.3 (6527) \\
95 & John Jaso & -7.76 (2.41) & [-12.50, -3.07] & 3172 & -11.3 (2879) \\ 
96 & Anthony Recker & -8.37 (2.33) & [-13.29, -3.93] & 2935 & -13 (3102) \\
97 & Gerald Laird & -8.68 (1.87) & [-12.29, -4.99] & 2378 & -9.6 (2616) \\ 
98 & A. J. Ellis & -12.90 (3.79) & [-20.10, -5.38] & 5476 & -12.3 (5345) \\
99 & Kurt Suzuki & -17.67 (4.25) & -[26.07, -9.35] & 6811 & -19.5 (7110) \\
100 & Dioner Navarro & -18.81 (4.68) & [-28.00, -9.40] & 6659 & -19.8 (6877)  \\
101 & Jarrod Saltalamacchia & -23.98 (4.35) & [-32.76, -15.87] & 6498 & -34 (6764) \\ \hline
\end{tabular}

\end{table}

According to our model, there is little posterior uncertainty that the framing effects of the top 10 catchers shown in Table~\ref{tab:runsSaved} had a positive impact for their teams, relative to the baseline catcher. 
Similarly, with the exception of Welington Castillo and Chris Iannetta, we are rather certain that the bottom 10 catchers' framing had an overall negative impact, relative to the baseline.
We estimate that Miguel Montero's framing saved his team 25.71 runs on average, relative to the baseline.
That is, had he been replaced by the baseline catcher on each of the 8,086 called pitches he received, his team would have given up an additional 25.71 runs, on average. 
Unsurprisingly, our estimates of framing impact differ from those of \citet{JudgePavlidisBrooks2015}'s model.
This is largely due to differences in the model construction, valuation of a called strike, and collection of pitches analyzed.
Indeed, in some cases, (e.g. Montero and Rene Rivera), they used more pitches to arrive at their estimates of runs saved while in others, we used more pitches (e.g. Mike Zunino and Jonathan Lucroy). 
Nevertheless, our estimate are not wholly incompatible with theirs; the correlation between our estimates and theirs is 0.94.  
Moreover, if we re-scale their estimates to the same number of pitches we consider, we find overwhelmingly that these re-scaled estimates fall within our 95\% posterior credible intervals.

\subsection{Catcher Aggregate Framing Effect}
\label{sec:safe2}

Looking at Table~\ref{tab:runsSaved}, it is tempting to say that  Miguel Montero is the best framer.
After all, he is estimated to have saved the most expected runs relative to the baseline catcher.
We observe, however, that Montero received 8086 called pitches while Conger received only 4743.
How much of the difference in the estimated number of runs saved is due to their framing ability and how much to the disparity in the called pitches they received?

A naive solution is to re-scale the RS estimates and compare the average number of runs saved on a per-pitch basis.
While this accounts for the differences in number of pitches received, it does not address the fact that Montero appeared with different players than Conger and that the spatial distribution of pitches he received is not identical to that of Conger. 
In other words, even if we convert the results of Table~\ref{tab:runsSaved} to a per-pitch basis, the results would still be confounded by pitch location, count, and pitch participants.

To overcome this dependence, we propose to \textit{integrate} $f(ca,\xi)$ over all $\xi$ rather than summing $f(ca,\xi)$ over $\mathcal{P}_{ca}.$
Such a calculation is similar to the spatially aggregate fielding evaluation (SAFE) of \citet{JensenShirleyWyner2009}.
They integrated the average number of runs saved by a player successfully fielding a ball put in play against the estimated density of location and velocity of these balls to derive an overall fielding metric un-confounded by dispraise in players' fielding opportunities.
%In that paper, they first estimated the average number of runs saved by a player successfully fielding a ball put in play as a function of the location and velocity of the ball. 
%They then integrated this against the estimated density of the location and velocity of balls put in play.
%The result was an overall fielding metric un-confounded by the disparities in players' fielding opportunities. 
%Unlike their setting, however, our confounding factors, $\xi$, are high-dimensional and, with the exception of the term related to pitch location, discrete. 
%This makes it difficult to construct a kernel density estimate of the distribution of $\xi$ in the same way that \citet{JensenShirleyWyner2009} did.
We propose to integrate $f(ca, \xi)$ against the empirical distribution of $\xi$ and define catcher $ca$'s ``Catcher Aggregate Framing Effect'' or CAFE to be 
\begin{equation}
\label{eq:safe2}
CAFE(ca) = 4000 \times \frac{1}{N}\sum_{\xi}{f(ca, \xi)},
\end{equation}

The sum in Equation~\ref{eq:safe2} may be viewed as the number of expected runs catcher $ca$ saves relative to the baseline if he participated in every pitch in our dataset.
We then re-scale this quantity to reflect the impact of his framing on 4000 ``average'' pitches. 
We opted to re-scale by CAFE by 4000 as the average number of called pitches received by catchers who appeared in more than 25 games was just over 3,992. 
Of course, we could have easily re-scaled by a different amount.

Once again, we can use our simulated posterior samples of the $\Theta^{u}$'s to simulate draws from the posterior distribution of CAFE. 
Table~\ref{tab:CAFE} shows the top and bottom 10 catchers ranked according to the posterior mean of their CAFE value, along with the posterior standard deviation, and 95\% credible interval for their CAFE value. 
Also shown is the a 95\% interval of each catchers marginal rank according to CAFE. 

\begin{table}[!h]
\centering
\footnotesize
\caption{Top and Bottom10 catchers according to the posterior mean of $CAFE.$}
\label{tab:CAFE}
\begin{tabular}{llccccc} \hline
Rank & Catcher & Mean (SD) & 95\% Interval & 95\% Rank Interval \\ \hline
1. & Hank Conger & 16.20 (2.72) & [10.84, 21.50] & [1, 11] \\
2. & Christian Vazquez & 14.33 (2.94) & [8.26, 20.03] &[1, 19] \\
3. & Rene Rivera & 14.04 (2.76) & [8.75, 19.31] & [1, 18] \\
4. & Martin Maldonado & 13.24 (3.33) & [6.73, 19.68] & [1, 24] \\
5. & Miguel Montero & 12.36 (2.42) &[7.50, 16.90]  & [2, 22] \\
6. & Yasmani Grandal & 11.90 (2.76) & [6.56, 17.29] & [2, 27] \\
7. & Mike Zunino & 11.78 (2.69) & [6.51, 16.74] & [2, 26] \\
8. & Chris Stewart & 11.63 (3.28)  & [5.21, 18.03] & [1, 30] \\
9. & Buster Posey & 11.16 (2.73) & [5.74, 16.51] & [2, 30] \\
10. & Francisco Cervelli & 10.45 (3.21) & [4.06, 16.72] & [2, 36] \\ \hline
92. & Jordan Pacheco & -11.73 (3.80) & [-19.26, -4.30]  & [68, 98] \\
93. & Koyie Hill & -11.79 (5.67)& [-22.48, -0.68] & [53, 100] \\
94. & Josh Phegley & -12.05 (4.66) & [-21.40, -3.20] & [64, 99] \\
95. & Austin Romine & -12.76 (9.78) & [-32.14, 5.81] & [30, 101] \\
96. & Jarrod Saltalamacchia & -14.00 (2.53) & [-19.11, -9.26] & [82, 99] \\
97. & Brett Hayes & -14.04 (4.06) & [-21.51, -5.93] & [73, 100] \\
98. & Gerald Laird & -14.96 (3.21) & [-21.17, -8.69] & [81, 99] \\
99. & Josmil Pinto & -15.04 (3.27) & [-21.57, -8.78] & [82, 100] \\
100. & Carlos Santana & -22.48 (4.63) & [-31.63, -13.26] & [93, 101] \\
101. &Tomas Telis & -25.06 (3.85) & [-32.41, -17.27] & [98, 101] \\ \hline
\end{tabular} 
\end{table}

We see that several of the catchers from Table~\ref{tab:runsSaved} also appear in Table~\ref{tab:CAFE}.
The new additions to the top ten, Christian Vazquez, Martin Maldonado, Chris Stewart, and Francisco Cervelli were ranked $13^{th}, 17^{th}, 18^{th}$ and $19^{th}$ according to the RS metric.
The fact that they rose so much in the rankings when we integrated over all $\xi$ indicates that their original rankings were driven primarily by the fact that they all received considerably fewer pitches in the 2014 season than the top 10 catchers in Table~\ref{tab:runsSaved}.
In particular, Vazquez received 3198 called pitches, Cervelli received 2424, Stewart received 2370, and Maldonado received only 1861.

Interestingly, we see that now Hank Conger ranks ahead of Miguel Montero according to the posterior mean CAFE, indicating that the relative rankings in Table~\ref{tab:runsSaved} was driven at least partially by disparities in the pitches the two received than by differences in their framing effects.
Though Conger emerges as a slightly better framer than Montero in terms of CAFE, the difference between the two is small, as evidenced by the considerable overlap in their 95\% posterior credible intervals. 

We find that in 95\% of the posterior samples, Conger had anywhere between the largest and $11^{th}$ largest CAFE.
In contrast, we see that in 95\% of our posterior samples, Tomas Tellis's CAFE was among the bottom 3 CAFE values. 
Interestingly, we find much wider credible intervals for the marginal ranks among the bottom 10 catchers.
Some catchers like Koyie Hill and Austin Romine appeared very infrequently in our dataset.
To wit, Hill received only 409 called pitches and Romine received only 61.
As we might expect, there is considerable uncertainty in our estimate about their framing impact, as indicated by the rather wide credible intervals of their marginal rank. 

\subsection{Year-to-year reliability of CAFE}
\label{sec:safe2_reliability}

We now consider how consistent CAFE is over multiple seasons.
We re-fit our model using data from the 2012 to 2015 seasons.
For each season, we restrict attention to those pitches within one foot of the approximate rule book strike zone from that season.
We also use the log-odds from the GAM models trained on all previous seasons so that the model fit to the 2012 data uses GAM forecasts trained only on data from 2011 while the model fit to the 2015 data uses GAM forecasts trained only on data from 2011 to 2014. 
When computing the values of CAFE, we use the run values given in Table~\ref{tab:run_values} for each season.
There were a total of 56 catchers who appeared in all four of these seasons.
Table~\ref{tab:year_to_year_CAFE} shows the correlation between their CAFE values over time.

\begin{table}[!h]
\centering
\caption{Correlation of CAFE across multiple seasons.}
\label{tab:year_to_year_CAFE}

\begin{tabular}{ccccc} \hline
~ & 2012 & 2013 & 2014 & 2015 \\ \hline
2012 & 1.00 & 0.70 & 0.56 & 0.41 \\
2013 & 0.70 & 1.00 & 0.71 & 0.61 \\
2014 & 0.56 & 0.71 & 1.00 & 0.58\\
2015 & 0.41 & 0.61 & 0.58 & 1.00 \\ \hline
\end{tabular}
\end{table}

In light of the non-stationarity in strike zone enforcement across seasons, it is encouraging to find moderate to high correlation between a player's CAFE in one season and the next.
In terms of year-to-year reliability, the autocorrelations of 0.5 -- 0.7 place CAFE on par with slugging percentage for batters. 
Interestingly, the correlations between 2012 CAFE and 2013 CAFE and the correlation between 2013 CAFE and 2014 CAFE are greater than 0.7, but the correlation between 2014 CAFE and 2015 CAFE is somewhat lower, 0.58.
While this could just be an artifact of noise, we do note that there was a marked uptick in awareness of framing between the 2014 and 2015 seasons, especially among fans and in the popular press.
One possible reason for the drop in correlation might be umpires responding to certain catcher's reputations as elite pitch framers by calling stricter strike zones, a possibility suggested by \citet{Sullivan2016}.

\section{Discussion}
\label{sec:discussion}

We systematically fit models of increasing complexity to estimate the effect a catcher has on an umpire's likelihood of calling a strike over and above factors like the count, pitch location, and other pitch participants.
We found evidence that some catchers do exert a substantially positive or negative effect on the umpires but that the magnitude of these effects are about as large as the count effects.
Using the model that best balanced fit and generalization, we were able to simulate draws from the posterior predictive distribution of the called strike probability probability of each taken pitch in 2014.
%We find that there is rather considerable uncertainty in the posterior predictive distribution.
For each pitch, we estimated the apparent framing effect of the catcher involved and, following a procedure similar to that of \citet{JudgePavlidisBrooks2015}, we derived an estimate of the impact framing has on the game, RS.
Our RS metric is largely consistent with previously reported estimates of the impact of catcher framing, but a distinct advantage is our natural quantification of the estimation uncertainty.
We find that there is considerable posterior uncertainty in this metric, making it difficult to estimate precisely the impact a particular catcher's framing had on his team's success.

While the construction of RS is intuitive, we argue that it does not facilitate reasonable comparisons of catchers' framing since, by construction, the metric is confounded by the other factors in our model.
We propose a new metric CAFE that integrates out the dependence of RS on factors like pitch location, count, and other pitch participants.
CAFE compares catchers by computing the impact each catcher's framing would have had had he received every pitch in our dataset.
Like RS, there is a considerable uncertainty in our CAFE estimates.
While we are able to separate the posterior distributions of CAFE of good framers from bad framers, there is considerable overlap in the posterior distributions of CAFE within these groups, making it difficult to distinguish between the good framers or between the bad framers.
Despite this, we find rather high year-to-year correlation in CAFE, though there is a marked drop-off between 2014 and 2015.
This coincides with the increased attention on framing in the sports media and sabermetrics community following the 2014 season.
One potential explanation for this drop-off is that umpires adjusted their strike zone enforcement when calling pitches caught by catchers with reputations as good framers. 

Our findings may have several implications for Major League Baseball teams.
The uncertainty in both RS and CAFE make it difficult to precisely value pitch framing with any reasonable degree of certainty. 
% Pick on Lucroy here
For instance, the 95\% credible interval of Jonathan Lucroy's RS is [8.16, 30.49].
Using the heuristic of 10 expected runs per win and $\$7$M per win \citep{Cameron2014, Pollis2013}, our model suggests that Lucroy's framing was worth anywhere between $\$5.7$M and $\$21.34$M. 
%For instance, the 95\% credible interval of Hank Conger's RS is $[13.187, 26.306]$
%Using the heuristic of 10 expected runs per win and $\$7$M per win, our model indicates that Conger's framing was worth anywhere between \$9.23M and \$18.4M.
In light of the non-stationarity between seasons and the recent drop-off in correlation in CAFE, it is difficult to forecast the impact that any individual catcher's framing will have into the future.
The observed overlaps in the posterior distribution of CAFE means that with a single season's worth of data, we cannot discriminate between good framers with the same certainty that we can separate good framers from bad framers.
As a concrete example, our model indicates that both Miguel Montero and Hank Conger were certainly better framers than Jarrod Saltalamacchia, but it cannot tell which of Montero or Conger had a larger, positive impact. 
%From a team's perspective, our results indicate that while there are catchers who can significantly influence umpires' decisions, it is difficult to estimate and value the impact this influence has over the course of a season with certainty. 

\subsection{Extensions}

% Add a bit saying that "we have not set out to provide a definitive model of pitch framing. Rather, our modeling setup is rather modular. There are many ways to improve the called strike probability forecasts. For instance, we could include factors like home/away. When including more of these factors, it may be fruitful to move to a more non-parametric approach, such as BART. 

% Another thing unexplored in this work: 

There are several extensions of and improvements to our model that we now discuss.
While we have not done so here, one may derive analogous estimates of RS and CAFE for batters and pitchers in a straightforward manner.
Our model only considered the count into which a pitch was thrown but there is much more contextual information that we could have included.
For instance, \cite{RosalesSpratt2015} have suggested the distance between where a catcher actually receives the pitch and where he sets up his glove before the pitch is thrown could influence an umpire's ball-strike decision making.
Such glove tracking data is proprietary but should it be become publicly available, one could include this distance along with its interaction with the catcher indicator into our model.
In addition, one could extend our model to include additional game-state information such as the ball park, the number of outs in the half-inning, the configuration of the base-runners, whether or not the home team is batting, and the number of pitches thrown so far in the at-bat. 
One may argue that umpires tend to call more strikes late in games which are virtually decided (e.g. when the home team leads by 10 runs in the top of the ninth inning) and easily include measures related to the run-differential and time remaining into our model.
Expanding our model in these directions may improve the overall predictive performance slightly without dramatically increasing the computational overhead.

More substantively, we have treated the umpires' calls as independent events throughout this paper. 
\citet{ChenMoskowitzShue2016} reported a negative correlation in consecutive calls, after adjusting for location.
To account for this negative correlation in consecutive calls, we could augment our model with binary predictors encoding the results of the umpires' previous $k$ calls in the same at-bat, inning, or game.
Incorporating this Markov structure to our model would almost certainly improve the overall estimation of called strike probability and may produce slightly smaller estimates of RS and CAFE. 
At this point, however, it is not \textit{a priori} obvious how large the differences would be or how best to pick $k.$
It is also well-known that pitchers try to throw to different locations based on the count, but we make no attempt to model or exploit this phenomenon.
Understanding the effect of pitch sequencing on umpires' decision making (and vice-versa) would also be an interesting line of future research. 

%One possible extension would be to fit a separate historical GAM for each combination of batter handedness, pitcher handedness, and count, though care must be taken to ensure sufficient sample size in each bin.

%In principle, it is possible to extend our model fitting strategy to encompass the fully parametrized model of Equation~\ref{eq:general_model}.
%The results presented here suggest that doing so will result in substantial overfitting, without extremely careful prior specification.
%In particular, the first level prior on the umpire-specific catcher effects would need to be very tightly concentrated around some unknown location to induce enough shrinkage. 
%It may also be the case that there exists a subset of catchers whose effect on a subset of umpires may be identified and it would be interesting to consider the question of pitch framing from a variable selection standpoint. 

We incorporated pitch location in a two-step procedure: we started from an already quite good generalized additive model trained with historical data and used the forecasted log-odds of a called strike as a predictor in our logistic regression model.
Much more elegant would have been to fit a single semi-parametric model by placing, say, a common Gaussian process prior on the umpire-specific functions of pitch location, $f^{u}(x,z)$ in Equation~\ref{eq:general_model}.
%Designing a computationally efficient MCMC procedure for such a model is a key challenge.
%\cite{DuranteDunson2014} derived a Gibbs sampler to fit logistic regression models of dynamic networks with Gaussian process priors using \citet{PolsonScottWindle2013}'s Polya-Gamma data augmentation strategy.
%A similar augmentation could be used here with suitable modifications made for the hierarchical structure.
%Approximation techniques like those described in \cite{BanerjeeDunsonTokdar2012} may also mitigate the computational burden of this semi-parametric approach. 
%Choosing an appropriate mean function and covariance kernel for the Gaussian process prior is also highly non-trivial.
We have also not investigated any potential interactions between pitch location, player, and count effects.
While we could certainly add interaction terms to the logistic models considered above, doing so vastly increases the number of parameters and may require more thoughtful prior regularization.
A more elegant alternative would be to fit a Bayesian ``sum-of-trees'' model using \citet{ChipmanGeorgeMcCulloch2010}'s BART procedure.
Such a model would likely result in more accurate called strike probabilities as it naturally incorporates interaction structure.
We suspect that this approach might reveal certain locations and counts in which framing is most manifest. 

Finally, we return to the two pitches from the 2015 American League Wild Card game in Figure~\ref{fig:keuchel_tanaka}.
Fitting our model to the 2015 data, we find that Eric Cooper was indeed much more likely to call the Keuchel pitch a strike than the Tanaka pitch (81.72\% vs 62.59\%).
Interestingly, the forecasts from the hGAMs underpinning our model were 51.31\% and 50.29\%, respectively.
%Cooper deviated substantially from the historical called strike probability estimates. 
Looking a bit further, had both catchers been replaced by the baseline catcher, our model estimates a called strike probability of 77.58\% for the Keuchel pitch and 61.29\% for the Tanaka pitch, indicating that Astros' catcher Jason Castro's apparent framing effect (4.14\%) was slightly larger than Yankee's catcher Brian McCann's (1.30\%). 
The rather large discrepancy between the apparent framing effects and the estimated called strike probabilities reveals that we cannot immediately attribute the difference in calls on these pitches solely to differences in the framing abilities of the catchers.
Indeed, we note that the two pitches were thrown in different counts: Keuchel's pitch was thrown in a 1 -- 0 count and Tanaka's was thrown in a 1 -- 1 count.
In 2015, umpires were much more likely to call strikes in a 1 -- 0 count than they were in a 1 -- 1 count, all else being equal.
Interestingly, had the Keuchel and Tanaka pitches been thrown in the same count, our model still estimates that Cooper would be consistently more likely to call the Keuchel pitch a strike, lending some credence to disappointed Yankees' fans' claims that his strike zone enforcement favored the Astros. 
Ultimately, though, it is not so clear that the differences in calls on the two pitches shown in Figure~\ref{fig:keuchel_tanaka} specifically was driven by catcher framing as much as it was driven by random chance. 

\newpage
\bibliography{DeshpandeWyner}
\newpage
\appendix

\section{Model Comparison with Cross-Validation}
\label{app:holdout}

As mentioned in Section~\ref{sec:model_comparison}, \citet{Roegelle2014}, \citet{Mills2017a}, and \citet{Mills2017b} have documented year-to-year changes in umpires' strike zone enforcement ever since Major League Baseball began reviewing and grading umpires' decisions in 2009.
In other words, umpire tendencies are non-stationary across seasons and we cannot reasonably expect Models 4 and 5, which attempt to identify umpire-specific player effects, to forecast future umpire decisions particularly well.
A potentially more appropriate way to diagnose overfitting issues would be to hold out a random subset of our 2014 data, say 10\%, fit each model on the remaining 90\% of the data, and assess the predictive performance on the held-out 10\%.
Table~\ref{tab:holdout_error} shows the average misclassification rate and mean square error for Models 1 -- 5 over 10 such holdout sets.
The results in the table confirm our finding that Model 3 represents the best balance between model expressivity and predictive capability.

\begin{table}[!h]
\centering

\caption{Hold-out misclassification rate (MISS) and mean square error (MSE) for several models)}
\label{tab:holdout_error}
\footnotesize
\begin{tabular}{llccccc} \\ \hline
~ & ~ Model 1 & Model 2 & Model 3 & Model 4 & Model 5 \\ 
~ & $\#$ Parameters & 744 & 855 & 2582 & 11,067 & 171,168  \\ \hline
Overall & MISS & 0.104 & 0.101 & \bf0.101 & 0.107 & 0.101 \\
~ & MSE & 0.073 & 0.071 & \bf 0.071 & 0.072 & 0.072 \\
Region 1 & MISS & 0.251 & \bf 0.239 & 0.240 & 0.242 & 0.240 \\
~ & MSE & 0.164 & 0.158 & \bf 0.157 & 0.159 & 0.158 \\
Region 2 & MISS & 0.213 & 0.208 & \bf0.206 & 0.208 & 0.207 \\
~ & MSE & 0.154 & 0.150 & \bf0.148 & 0.150 & 0.149 \\ \hline
\end{tabular}
\end{table}

\section{Catcher and Count Effects}
\label{app:catcher_count_effects}

In Section~\ref{sec:full_posterior_analysis}, we reported that the posterior distributions of catcher and count effects on the log-odds scale were largely supported in the interval [-1.5, 1.5].
This would indicate that a catcher's framing effect is of roughly similar magnitude as the effect of count.

\begin{figure}[!h]
\centering
\includegraphics{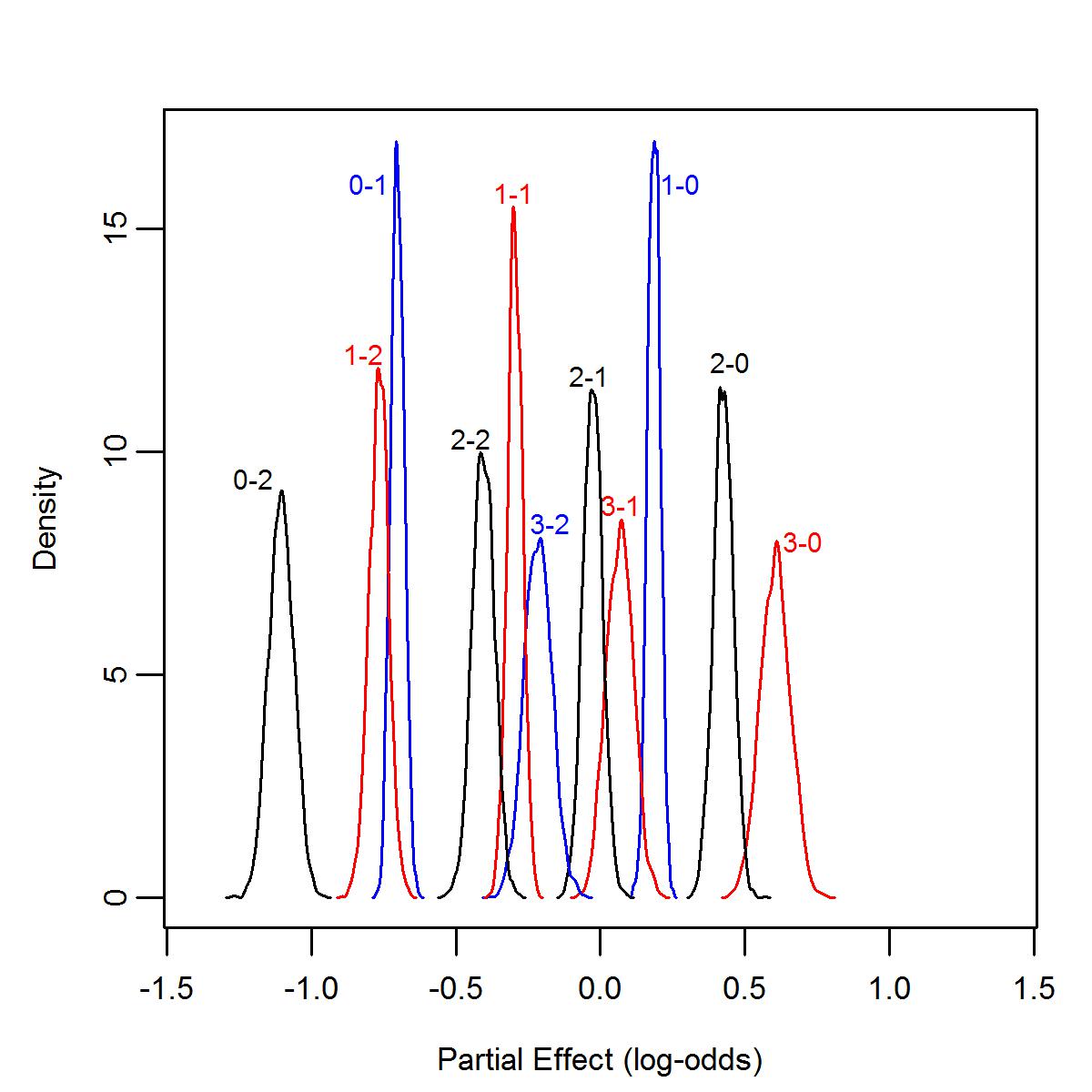}
\caption{Posterior densities of the partial effect of count. Densities computed using a standard kernel estimator}
\label{fig:count_densities}
\end{figure}

Figure~\ref{fig:count_densities} shows the approximate posterior densities of the count effects.
Recall that these are the partial effects relative to the baseline count of 0 -- 0.
As we might expect, umpires are more likely to call strikes in 3 -- 0 and 2 -- 0 counts than in 0 -- 0 counts and much less likely to call strikes in 0 -- 2 and 1 -- 2 counts, all else being equal.
Somewhat interestingly, we find that umpires are slightly less likely to call strikes in a 3 -- 1 count than they are in a 1 -- 0 count.

We also see that the posterior distribution for the effects of a 3 -- 0  counts are considerably wider than those for a 0 -- 1 count, indicating that we are much more uncertain about the effect of the former two counts than the latter two.
This is due the rather large disparity in the numbers of pitches taken in these counts: in our dataset, there were more than five times called pitches thrown into a 0 -- 1 count than into a 3 -- 0 count (37,513 versus 6,162). 

To compare the relative magnitudes of catcher and count effects on the probability scale, we return to the hypothetical matchup between batter Yasiel Puig, catcher Buster Posey, and pitcher Madison Bumgarner. 
Suppose that Bumgarner's pitch is thrown in a location where the hGAM called strike probability forecast is exactly 50\%.
According to our model, if this pitch were hypothetically caught by the baseline catcher, Brayan Pena, the forecasted called strike probability averaged over all 93 umpires is 54\%, with the difference of 4\% attributable to intercept, batter, and pitcher effects.  
In contrast, if the same pitch had been caught by Buster Posey, our model estimates the called strike probability to be 64\%, indicating that on this pitch, Posey added an additional 10\% to the forecasted called strike probability. 
If Posey had caught the same pitch but the count were 2 -- 2 instead of 0 -- 0, the forecast would be 55\%.
In this way, at least for this pitch, the effect of swapping the baseline catcher with Posey on an 0 -- 0 pitch is the about the same as changing the count from 0 -- 0 to 2 -- 2 with Posey catching.

Table~\ref{tab:catcher_count_effect_size} elaborates on this example and shows the estimated average called strike probability for the same pitch as a function of count and catcher.
That is, we forecasted the called strike probability, averaged across umpires, on a pitch thrown by Bumgarner to Puig at a location where the hGAM forecast was 50\% for many combinations of catcher and count. 
To highlight the relative size of the catcher and count effects, we have subtracted a baseline 54\%, the called strike probability when the catcher is Pena and the count is 0 -- 0, from all of these probabilities.

\begin{table}[!h]
\centering
\caption{Difference in forecasted called strike probabilities averaged over all umpires  when Bumgarner pitches to Puig, relative to the baseline called strike probability of 54\%, for various combinations of catcher and count. Counts are ordered as in Figure~\ref{fig:count_densities}}
\label{tab:catcher_count_effect_size}
\footnotesize

\begin{tabular}{lcccc} \\ \hline
~ &Hank Conger & Buster Posey & Brayan Pena & Tomas Telis \\ \hline
0 -- 2 & -0.14 & -0.18 & -0.26 & -0.39 \\
1 -- 2 & -0.06 & -0.10 & -0.18 & -0.35 \\
0 -- 1 & -0.04 & -0.08 & -0.17 & -0.34 \\
2 -- 2 & 0.03 & -0.01 & -0.10 & -0.28 \\
1 -- 1 & 0.06 & 0.02 & -0.07 & -0.26 \\
3 -- 2 & 0.07 & 0.0 4& -0.05 & -0.25 \\ 
2 -- 1 & 0.12 & 0.08 & -0.01 & -0.20 \\
0 -- 0 & 0.12 & 0.09 & 0.00 & -0.20 \\
3 -- 1 & 0.14 & 0.10 & 0.02 & -0.18 \\
1 -- 0 & 0.16 & 0.13 & 0.04 & -0.16 \\
2 -- 0 & 0.21 & 0.18 & 0.10 & -0.10 \\
3 -- 0 & 0.24 & 0.21 & 0.14 & -0.06 \\ \hline
\end{tabular}
\end{table}

Recall that the baseline called strike probability is 54\% on such a pitch. 
According to our model, the effect of changing the count from 0 -- 0 to 0 -- 2 when Posey is receiving the pitch is about the same as changing the count from 0 -- 0 to 1 -- 2 with the baseline catcher receiving the pitch. 
We note that the called strike probability forecasts for Tomas Tellis are much lower than for Hank Conger, Posey, and the baseline catcher. 
For instance, our model estimates that umpires on average would call this pitch a strike only 15\% of the time if it were thrown in a 0 -- 2 count and Tellis was receiving, in contrast to 40\% for Conger, 36\% for Posey, and 28\% for Pena. 

\end{document}